\documentclass[a4paper,11pt]{article}
\pdfoutput=1 
\usepackage{jheppub} 

\usepackage[T1]{fontenc} 

\title{\boldmath Flavor origin of dark matter and its relation with leptonic nonzero $\theta_{13}$
           and Dirac CP  phase $\delta$ }

\author[a]{Subhaditya Bhattacharya}
\author[a]{Biswajit Karmakar,}
\author[b]{Narendra Sahu}
\author[a]{Arunansu Sil}

\affiliation[a]{Department of Physics, Indian Institute of Technology Guwahati, 781039 Assam, India}
\affiliation[b]{Department of Physics, Indian Institute of Technology,
 Hyderabad, Kandi, Sangareddy 502285, Medak, Telengana, India}

\emailAdd{subhab@iitg.ernet.in}
\emailAdd{k.biswajit@iitg.ernet.in}
\emailAdd{nsahu@iith.ac.in}
\emailAdd{asil@iitg.ernet.in}

\arxivnumber{1611.07419}

\abstract{We propose a minimal extension of the standard model by including a $U(1)$ flavor symmetry to establish a 
correlation between the relic abundance of dark matter, measured by WMAP and PLANCK satellite experiments 
and non-zero value of $\sin \theta_{13}$ observed at DOUBLE CHOOZ, Daya Bay, RENO and T2K. The flavour 
symmetry is allowed to be broken at a high scale to a remnant $\mathcal{Z}_2$ symmetry, which not only ensures the 
stability to the dark matter, but also gives rise to a modification to the existing $A_4$-based tri-bimaximal 
neutrino mixing. This deviation in turn suggests the required non-zero value of $\sin \theta_{13}$. We assume 
the dark matter to be neutral under the existing $A_4$ symmetry while charged under 
the $U(1)$ flavor symmetry. Hence in this set-up, the non-zero value of $\sin \theta_{13}$ predicts the dark matter charge 
under $U(1)$, which can be tested at various ongoing and future direct and collider dark matter search experiments. 
We also point out the involvement of nonzero leptonic CP phase $\delta$, which plays an important role in the 
analysis.}

\begin{document} 
\maketitle
\flushbottom

\section{Introduction}
After the Higgs discovery at the LHC, the standard model (SM) of particle physics seems to be complete. 
However, it does not explain many current issues in particle physics which are supported by experiments. 
In particular, the oscillation experiments~\cite{sk1,sk2,MINOS,KamLand} confirm that the neutrinos are massive and they mix 
with each other. Contrary to this finding, neutrinos are massless within the framework of SM. Another outstanding 
problem in particle physics as of today is the nature of dark matter (DM), whose relic abundance is precisely 
measured by the WMAP~\cite{wmap} and PLANCK~\cite{planck} satellite experiments to be $0.094 \leq \Omega_{\rm DM}h^2 
\leq  0.130$. In fact, the existence of DM is strongly supported by the galactic rotation curve, gravitational 
lensing and large scale structure of the Universe~\cite{DM_review} as well. However, the SM of particle physics 
fails to provide a candidate of DM. In this work our aim is to go beyond the SM of particle physics to explore 
scenarios which can accommodate a candidate of DM as well as non-zero neutrino masses and mixings.  

Flavor symmetries are often used to explore many unsolved issues within and beyond the SM of particle physics. 
For example, a global $U(1)$ flavor symmetry was proposed a long ago to explain the quark mass hierarchy and 
Cabibbo mixing angle~\cite{Froggatt:1978nt}. Subsequently many flavor symmetric frameworks have been adopted to 
explain neutrino masses and mixings in the lepton sector. In particular, a tri-bimaximal (TBM) lepton mixing 
generated from a discrete flavor symmetry such as $A_4$ attracts a lot of attention~\cite{Ma:2001dn,Altarelli:2005yx} 
due to its simplicity and predictive nature. However the main drawback of these analyses was that it predicts vanishing 
reactor mixing angle $\theta_{13}$ which is against the recent robust observation of $\theta_{13} \approx 
9^{\circ}$~\cite{Capozzi:2013csa,Gonzalez-Garcia:2014bfa, Forero:2014bxa} by DOUBLE CHOOZ~\cite{Abe:2011fz}, 
Daya Bay~\cite{An:2012eh}, RENO~\cite{Ahn:2012nd} and T2K~\cite{Abe:2013hdq} experiments. Hence, a modification 
of the TBM structure of lepton mixing is required. 

In this work we consider the existence of a dark sector ~\cite{Bhattacharya:2015qpa} consisting of vector-like 
fermions which are charged under an additional $U(1)$ flavor symmetry. Specifically, we consider a vector-like SM
singlet fermion ($\chi^0$) and a $SU(2)_L$ doublet fermion ($\psi$) which are odd under the remnant $\mathcal{Z}_2$ symmetry 
generated from the broken $U(1)$. The neutral components mix to give rise a fermionic DM ($\psi_1$). Note that in 
the simplest case, a singlet fermion ($\chi^0$) can generate a Higgs portal interaction by dimension five operator 
suppressed by the new physics scale as $(\overline{\chi^0}\chi^0 H^{\dagger}H)/\Lambda$. However, as we argue, that 
the new physics scale ($\Lambda$) involved in the theory has to generate the required neutrino mass as well and thus 
making it very high. As a result, the annihilation rate of DM becomes too small which in turn make the relic density over 
abundant. On the other hand, a vector-like fermion doublet ($\psi$) suffers from a large annihilation cross-section 
to SM through $Z$ mediation and is never enough to produce the required density. It is only through the mixing of these 
two that can produce correct relic density as we demonstrate here. We also assume the existence of a TBM neutrino 
mixing pattern (in a basis where charged leptons are diagonal) based on $A_4$ symmetry. The interaction between the dark 
and the lepton sector of the SM is mediated by flavon fields charged under the $U(1)$ and/or $A_4$. These flavons also 
take part in producing additional interactions involving lepton and Higgs doublets. The $U(1)$ symmetry, once allowed 
to be broken by the vacuum expectation value (vev) of a flavon, generates a non-zero $\sin \theta_{13}$ after the 
electroweak symmetry breaking (and when $A_4$ breaks too). We show that the non-zero value of $\sin \theta_{13}$ is proportional 
to the strength of Higgs portal coupling of DM giving rise to the correct relic density. In other words, the precise value of 
$\sin \theta_{13}$ and DM relic density can fix the charge of dark matter under $U(1)$ flavor symmetry. Indeed it 
is true for the Dirac CP violating phase $\delta=0$ as shown in our previous work~\cite{Bhattacharya:2016lts}.
However, we have found here that the non-zero values of $\delta$ plays an important role for the determination of DM 
charge under $U(1)$ flavor symmetry. Although 
the current allowed range of $\delta$ ($ 0^\circ - 360^\circ $) can significantly increase the uncertainty 
in the determination of DM flavor charge (compared to $\delta=0$ scenario), a future measurement of $\delta$ would be important in 
fixing the charge. In~\cite{Bhattacharya:2016lts}, we have assumed a prevailing TBM pattern and here in this work we 
provide an explicit construction of that too.  We also show that the effective Higgs portal coupling of the vector-like 
leptonic DM can be tested at future direct search experiments, such as Xenon1T~\cite{Aprile:2015uzo} and at the Large Hadron 
Collider (LHC)~\cite{Bhattacharya:2015qpa,Arina:2012aj,Arina2}. 

The draft is arranged as follows. In section \ref{model_part} we discuss the relevant model for correlating non-zero 
$\sin \theta_{13}$ to Higgs portal coupling of DM which gives correct relic density. In section \ref{neutrino_section}
and \ref{DM_section}, we obtain the constraints 
on model parameters from neutrino masses and mixing and relic abundance of dark matter respectively. In section \ref{DM_Neutrino}, we obtain 
the correlation between the non-zero $\sin \theta_{13}$ and Higgs portal coupling of dark matter and conclude in section \ref{concl}.   

\section{Structure of the model} \label{model_part}
In this section, we describe the field content and symmetries involved. We consider an effective field theory approach 
for realizing the neutrino masses and mixing while trying to connect it with the DM sector as well. The set-up includes 
the interaction between these two sectors which has the potential to generate adequate $\theta_{13}$, and hence a deviation of TBM mixing happens, 
to match with the experimental observation while satisfying the constraints from relic density and direct search of DM. 

\subsection{Neutrino Sector}

\begin{table}[h]
\centering
\resizebox{12cm}{!}{%
\begin{tabular}{|c|ccccc|cc|ccccc|}
\hline
 Field & $e_{R}$ & $\mu_{R}$ & $\tau_{R}$  & $\ell$ & $H$ & $\psi$ & 
 $\chi^o$ & $\phi_S$ & $\phi_T$ &  $\xi$ & $\eta$ & $\phi$ \\
\hline
$SU(2)_L$ &1 & 1 & 1 & 2 & 2 & 2 & 1 & 1 & 1 & 1 & 1 &1\\ 
\hline
$A_{4}$ & 1 & $1''$ & $1'$ &  3 & 1 & 1 & 1 & 3 & 3 & 1&$1'$ &1 \\
\hline
$Z_{3}$ &$\omega$ & $\omega$ & $\omega$ & $\omega$ & 1 & 1 &
1 & $\omega$ & 1 & $\omega$ & $\omega$ &1 \\
\hline
$Z_{2}$ &-1 & -1 & -1 & 1 & 1 & -1 & -1 & 1 & -1 & 1 & 1 &1 \\
\hline
$U(1)$  & 0 & 0 & 0  & 0 & 0& $q_1$ & $q_2$ & 0 & 0 & 0& $-x$ & $x$\\
\hline 
\end{tabular}
}\
\caption{\label{tab:1} {\small Fields content and transformation properties
under the symmetries imposed on the model. Here $nx = q_1 - q_2$ (justified from Eq.(\ref{lagrangian})), $n$ will be determined later.}}
\end{table}

The basic set-up relies on the $A_{4}$ symmetric construction of the Lagrangian associated with neutrino mass 
term~\cite{Ma:2001dn,Altarelli:2005yx}. Based on the construction by Altarelli-Feruglio (AF)
model~\cite{Altarelli:2005yx} (for generating TBM mixing), we have extended the flavon sector and symmetry of the model.  
The SM doublet leptons ($\ell$) transform as triplet under the $A_4$ symmetry 
while the singlet charged leptons: $e_R, \mu_R$ and $\tau_R$ transform as $1, 1^{''}$ and $1^{'}$ respectively under $A_4$. The 
flavon fields and their charges (along with the SM fields) are described in Table \ref{tab:1}. The flavons $\phi_S, \phi_T$ 
and $\xi$  break the $A_4$ flavor symmetry by acquiring vevs in suitable directions. Note that here $\phi_S$ and $ \phi_T$  
transform as $A_4$ triplets but the flavon $\xi$ and the SM Higgs doublet ($H$) transform as a singlet under $A_4$. 
So the contribution to  the effective neutrino mass matrix coming through the higher dimensional operator respecting 
the symmetries considered can be written as
\begin{equation}\label{lm0}
 -\mathcal{L}_{\nu_0}=(\ell H \ell H)(y_1\xi- y_2 \phi_S)/\Lambda^2\,,
\end{equation}
where $\Lambda$ is the cut off scale of the theory and $y_1,y_2$ represents respective coupling constant. The scalar fields break the 
flavor symmetry when acquire vevs along\footnote{The chosen vev alignments of $\phi_S$ and $\phi_T$ can be obtained by minimizing the
potential involving them along a similar line followed in ~\cite{Altarelli:2005yx, King:2011zj,Holthausen:2012wz,Dorame:2012zv}.}
$\langle \phi_S \rangle=(v_S,0,0)$, $\langle \phi_T \rangle=v_T(1,1,1)$, $\langle \xi \rangle=v_{\xi}$ 
and $\langle H \rangle=v$. As a result we obtain the light neutrino mass matrix as
\begin{eqnarray}\label{neutrino-mass}
 (m_{\nu})_0=\left(
\begin{array}{ccc}
      a-2b/3   &b/3    &b/3\\
       b/3    &-2b/3    &a+b/3\\
       b/3    &a+b/3   &-2b/3
\end{array}
\right)\,,
\end{eqnarray}
where $a=y_1(v^2/\Lambda)\epsilon$ and $b=y_2(v^2/\Lambda)\epsilon$, with $\epsilon=v_{\xi} / \Lambda=v_S / \Lambda$ is considered without 
loss of generality as any prefactor (due to the mismatch of vevs)  can be absorbed in the definition of $y_2$. 
The above mass matrix can be diagonalized by the TBM mixing matrix matrix~\cite{Harrison:1999cf}
 \begin{eqnarray}\label{utb}
 U_{TB}=\left(
\begin{array}{ccc}
 \sqrt{\frac{2}{3}} & \frac{1}{\sqrt{3}} & 0 \\
 -\frac{1}{\sqrt{6}} & \frac{1}{\sqrt{3}} & -\frac{1}{\sqrt{2}} \\
 -\frac{1}{\sqrt{6}} & \frac{1}{\sqrt{3}} & \frac{1}{\sqrt{2}}
\end{array}
\right). 
\end{eqnarray}
The relevant contribution to charged leptons (considering charges from Table \ref{tab:1}) can be obtained via 
\begin{equation}
 \mathcal{L}_l =  \frac{y_e}{\Lambda}(\bar{\ell}\phi_T)H e_R
+\frac{y_{\mu}}{\Lambda}(\bar{\ell}\phi_T)'H\mu_R+ 
\frac{y_{\tau}}{\Lambda}(\bar{\ell}\phi_T)''H\tau_R\,,
\end{equation} 
which yields the diagonal mass matrix: 
\begin{eqnarray}
M_{l} =  \left(
\begin{array}{ccc}
   y_{e}v\frac{v_T}{\Lambda} &0 & 0\\
   0    & y_{\mu}v\frac{v_T}{\Lambda}     & 0\\
       0 &0 &y_{\tau}v\frac{v_T}{\Lambda} 
\end{array}
\right).
\end{eqnarray}
Note that this is the leading order contribution (and is proportional to $1/\Lambda$) in the charged lepton mass matrix. Due to
the symmetry of the model as described in Table \ref{tab:1} (including the $U(1)$ symmetry to be discussed later) there will be
no term proportional to $1/\Lambda^2$. Therefore  no contribution to the lepton mixing matrix originated from the
charged lepton sector up to $1/\Lambda^2$ is present. Here it is worthy to mention that the dimension-5 operator $\ell H \ell H/\Lambda$ is 
forbidden due to the $Z_3$ symmetry specified in Table \ref{tab:1}. This additional symmetry also forbids the dimension-6 operator
$\ell H \ell H(\phi_T+\phi_T^{\dagger})/\Lambda^2$. The $U(1)$ flavor symmetry considered here does not allow terms involving
$\phi,\eta$ (such as: $\ell H \ell H(\phi+\eta)/\Lambda^2$) as 
discussed (where $\phi$ and $\eta$ are charged under $U(1)$ but the SM particles are not). Therefore, Eq. (\ref{lm0}) is the 
only relevant term up to $1/\Lambda^2$ order contributing to the neutrino mass matrix $(m_{\nu})_0$ ensuring its TBM structure as 
in Eq. (\ref{neutrino-mass}). Note that these kind of structure of the neutrino mass matrix of $(m_{\nu})_0$ can also be obtained 
in a $A_4$ based set-up either in a type-I, II or inverse seesaw framework~\cite{Karmakar:2014dva,Branco:2012vs,Karmakar:2015jza, Karmakar:2016cvb}. 

The immediate consequence of TBM mixing as given in Eq. (\ref{utb}) is that it implies  $\sin^2 \theta_{12}=1/3$ , $\sin^2 \theta_{23}=1/2$ 
and $\sin \theta_{13}=0$. Now to explain the current experimental observation on $\theta_{13}$ we consider an operator of 
order $1/\Lambda^3$:
\begin{equation}
  -\delta\mathcal{L}_{\nu}=y_3\frac{(\ell H \ell H)\phi\eta}{\Lambda^3}\,,
\end{equation}
where we have introduced two other SM singlet flavon fields $\phi$ and $\eta$ which carry equal and opposite charges under the $U(1)$ 
symmetry but transform as $1$ and $1^{'}$ under $A_4$ respectively. The $U(1)$ charge assignment to these two flavons also ensures that 
$\phi$ and $\eta$ do not take part in $(m_{\nu})_0$. Thus, after flavor and electroweak symmetry breaking this term contributes 
to the light neutrino mass matrix as follows: 
\begin{equation}\label{neutrino-correction}
\delta m_\nu= \left(
\begin{array}{ccc}
      0   &0   &d\\
      0    &d   &0\\
      d    &0  &0
\end{array}
\right), 
\end{equation} 
where $d=y_3(v^2/\Lambda)\epsilon^2$ with $\epsilon=\langle \phi \rangle/ \Lambda \equiv \langle \eta \rangle /\Lambda$.
This typical flavor structure of the additional contribution in the neutrino mass matrix follows from the involvement of $\eta$ field,
which transforms as $1'$ under $A_4$~\cite{Shimizu:2011xg, Karmakar:2014dva}. This $\delta m_\nu$ can indeed generate the 
$\theta_{13}\neq 0$ in the same line as in~\cite{Karmakar:2014dva,Karmakar:2015jza,Karmakar:2016cvb}. Note that the choice of $Z_2$
symmetry presented in  Table \ref{tab:1} also forbids the contributions to neutrino mass matrix proportional to $1/\Lambda^3$ 
(involving terms like $\ell H \ell H\phi_S\phi_T, \ell H \ell H\xi\phi_T, \ell H \ell H\phi_S\phi_T^{\dagger}$ and 
$\ell H \ell H\xi\phi_T^{\dagger}$) and thus 
ensuring Eq. (\ref{neutrino-correction}) is the only contribution responsible for breaking the  TBM mixing.

\subsection{Dark sector and its interaction with neutrino sector}

The dark sector associated with the present construction consists of a vector-like $SU(2)_L$ doublet $\psi^T=(\psi^0,\psi^-)$ and a 
neutral singlet fermion $\chi^0$ \cite{Bhattacharya:2015qpa}, which are odd under the $Z_2$ symmetry as has already been mentioned in 
Table \ref{tab:1}. These fermions are charged under an additional $U(1)$ flavor symmetry, but neutral under the existing 
symmetry in the neutrino sector (say the non-Abelian $A_4$ and additional discrete symmetries required).  Note that all the SM 
fields and the additional flavons in the neutrino sector except $\phi$ are neutral under this additional $U(1)$ symmetry. Since 
$\psi$ and $\chi^0$ are vector-like fermions, they can have  bare masses, $M_\psi$  and $M_\chi$, which are not protected by the SM symmetry. 
The effective Lagrangian, invariant under the symmetries considered, describing the interaction between the dark and the SM sector 
is then given by: 
\begin{equation}\label{lagrangian}
 \mathcal{L}_{\rm int}=\left(\frac{\phi}{\Lambda}\right)^n 
\overline{\psi}\widetilde{H}\chi^0,
\end{equation}
where $n$ is not fixed at this stage. The above term is allowed provided the $U(1)$ charge of $\phi^n$ is compensated by
$\psi$ and $\chi^0$ $i.e.$ $nx=q_1-q_2$. We will fix it later from phenomenological point of view.  

When $\phi$  acquires a vev, the $U(1)$ symmetry breaks down and an effective Yukawa interaction is generated 
between the SM and the DM sectors. After electroweak symmetry is broken, the DM emerges as an admixture of the
neutral component of the vector-like fermions $\psi$, {\it i.e.} $\psi^0$, and $\chi^0$. The Lagrangian describing 
the DM sector and the interaction as a whole reads as 
\begin{equation}\label{eq:Lag-lepton}
\hspace*{-0.5cm}
 -\mathcal{L_{\rm Yuk}}  \supset  M_\psi \overline{\psi}\psi + M_\chi \overline{\chi^0}\chi^0 + \left[ Y\overline{\psi}\widetilde{H}\chi^0 
+ {\rm h.c.}\right]\,,
\end{equation}
where the effective Yukawa connecting the dark sector to the SM Higgs reads as $Y=\epsilon^n=\left(\frac{\langle \phi \rangle}{\Lambda}\right)^n$.
We have already argued in introduction about our construction of dark matter sector. 
The idea of introducing vector-like fermions in the dark sector is also motivated by the fact that we expect a replication of the 
SM Yukawa type interaction to be present in the dark sector as well. Here the $\phi$ field plays the role of the messenger field 
similar to the one considered in~\cite{Calibbi:2015sfa}. See also ~\cite{Hirsch:2010ru,Boucenna:2011tj,deAdelhartToorop:2011ad,
Hamada:2014xha,Huang:2014jaa,Lattanzi:2014mia,Varzielas:2015joa,Ma:2015roa,Varzielas:2015sno,Mukherjee:2015axj,Lamprea:2016egz} 
for some earlier efforts to relate $A_4$ flavor 
symmetry to DM. Note that the vev of the $\phi$ field is also instrumental in producing the term $d$ to the neutrino mass matrix along 
with the vev of $\eta$. Since the $d$-term is responsible for generation of nonzero $\theta_{13}$ (will be discussed in the next 
section) a connection between non-zero $\sin\theta_{13}$ and DM interaction becomes correlated in our set-up. 

A discussion about other possible terms allowed by the symmetries considered would be pertinent here. Terms like 
$\overline{\psi}\psi H^{\dagger} H/{\Lambda}$ and $\overline{\chi^0}\chi^0 H^{\dagger} H/{\Lambda}$ are 
actually allowed in the present set-up. However it turns out that their role is less significant compared to the other
terms present. The reason is the following: firstly they could contribute to bare mass terms of $\psi$ and $\chi^0$ fields. 
However these contribution being proportional to $v^2/{\Lambda}$ are insignificant as compared to $M_\psi$ and $M_\chi$.
Similar conclusion holds for the Yukawa term as well. Secondly, they could take part in the DM annihilation. However as we 
will see, there also they do not have significant contribution because of the $\Lambda$ suppression.

\section{Phenomenology of the neutrino sector}\label{neutrino_section}
Combining Eqs. (\ref{neutrino-mass}) and (\ref{neutrino-correction}), we get the light neutrino mass 
matrix as $m_\nu=(m_\nu)_0 + \delta m_\nu$. We have already seen that $(m_\nu)_0$ can be 
diagonalized by $U_{TB}$ alone. Hence including $\delta m_{\nu}$, rotation by $U_{TB}$ results into 
the following structure of neutrino mass matrix:

\begin{eqnarray}
m'_{\nu}&=&U_{TB}^Tm_{\nu}U_{TB},\\
&=&\left(
\begin{array}{ccc}
 a-b-d/2     & 0   & \sqrt{3}d/2 \\
 0           & a+d & 0 \\
 \sqrt{3}d/2 & 0   & -a-b+d/2\\
\end{array}\label{mnupr}
\right).
\end{eqnarray}
So an additional rotation (by the $U_1$ matrix given below) is required to diagonalize $m_\nu$, 
\begin{equation}\label{mnudia}
(U_{TB}U_1)^Tm_{\nu}(U_{TB}U_1) = {\rm diag} (m_1e^{i\gamma_1},m_2e^{i\gamma_2}, 
m_3e^{i\gamma_3})\,
\end{equation}
where 
\begin{eqnarray}\label{u1}
U_1 =\left(
\begin{array}{ccc}
 \cos\theta_{\nu}                & 0 & \sin\theta_{\nu}e^{-i\varphi}  \\
     0                           & 1 &            0 \\
 -\sin\theta_{\nu}e^{i\varphi}  & 0 &        \cos\theta_{\nu} 
\end{array}
\right)\,. 
\end{eqnarray}
Here $m_{i=1,2,3}$ are the real and positive eigenvalues and $\gamma_{i=1,2,3}$ are the phases 
associated to these mass eigenvalues. We can therefore extract
the neutrino mixing matrix $U_{\nu}$ as, 
\begin{eqnarray}\label{unu}
 U_{\nu} =U_{TB}U_1U_m= \left(
\begin{array}{ccc}
 \sqrt{\frac{2}{3}}\cos\theta_{\nu} & \frac{1}{\sqrt{3}} & 
 \sqrt{\frac{2}{3}} e^{-i \varphi } \sin\theta_{\nu} \\
 -\frac{\cos\theta_{\nu}}{\sqrt{6}}+\frac{e^{i \varphi } \sin\theta_{\nu}}{\sqrt{2}}
 & \frac{1}{\sqrt{3}}
 & -\frac{\cos\theta_{\nu}}{\sqrt{2}}-\frac{e^{-i \varphi }\sin\theta_{\nu}}{\sqrt{6}} \\
 -\frac{\cos\theta_{\nu}}{\sqrt{6}}-\frac{e^{i \varphi } \sin\theta_{\nu}}{\sqrt{2}} & 
 \frac{1}{\sqrt{3}} & \frac{\cos\theta_{\nu}}{\sqrt{2}}-\frac{e^{-i \varphi }
 \sin\theta_{\nu}}{\sqrt{6}}
\end{array}
\right)U_{m}\,,
\end{eqnarray}
where $U_{m}={\rm diag} (1,e^{i\alpha_{21}/2},e^{i\alpha_{31}/2})$ is 
the Majorana phase matrix with $\alpha_{21}=(\gamma_1-\gamma_2)$ and 
$\alpha_{31}=(\gamma_1-\gamma_3)$, one common phase being irrelevant.
The angle $\theta_{\nu}$ and phase $\varphi$ associated in $U_1$ can now be linked
with the parameters: $a,b,d$ involved in $m_{\nu}$ through Eq. (\ref{mnupr}). 

Note that the parameters: $a,b$ and $d$ are
all in general complex quantities. 
We define the phases associated with $a,b, d$ as $\phi_a, \phi_b$ and $\phi_d$ respectively. 
Also for simplifying the analysis, we consider $|y_1|=|y_3|=y$ and $|y_2|=k$. With these, 
$\theta_{\nu}$ and $\varphi$ can be expressed in terms of the parameters involved in the 
effective light neutrino mass matrix $m'_{\nu}$ as:
\begin{eqnarray}\label{tpsi}
\tan{2\theta_{\nu}} &=&\frac{\sqrt{3}\epsilon\cos\phi_{db}}
                             {(\epsilon\cos\phi_{db}-2\cos\phi_{ab})\cos\varphi},\\          
\tan\varphi&=&\frac{y}{k}\frac{\sin(\phi_{db}-\phi_{ab})}
                 {\cos\phi_{db}}\label{tpsi2}.           
\end{eqnarray}
where $\phi_{ab}=\phi_a-\phi_b$ and $\phi_{db}=\phi_d-\phi_b$.
Then comparing the standard $U_{PMNS}$ parametrization and neutrino mixing matrix $U_{\nu}(=U_{TBM}U_1U_m)$ we obtain 
\begin{eqnarray}
\sin\theta_{13}=\sqrt{\frac{2}{3}}\left|\sin\theta_{\nu}\right|, \hspace{.3cm}
\delta={\rm arg}[(U_1)_{13}] 
\label{ang}. 
\end{eqnarray}
From Eq. (\ref{tpsi}) and (\ref{tpsi2}) it is clear that, $\sin\theta_{\nu}$  may take positive or negative value depending on the choices of
$\epsilon$ and $y/k$. For $\sin\theta_{\nu}>0$, we find $\delta=\varphi$ using $\delta={\rm arg}[(U_1)_{13}]$. On the other hand for 
$\sin\theta_{\nu}<0$; $\delta$ and $\varphi$ are related by $\delta=\varphi \pm \pi$. 
Therefore in both these cases we obtain $\tan{\varphi} = \tan{\delta}$ and hence Eq. (\ref{tpsi2}) leads to 
\begin{equation}
\tan\delta  = \frac{y}{k}\frac{\sin(\phi_{db}-\phi_{ab})}
                 {\cos\phi_{db}}.
\label{delta-phid}
\end{equation}
The other two mixing angles follow the standard correlation with $\theta_{13}$ in $A_4$ models~\cite{Altarelli:2010gt,King:2011zj}. 

Using Eq. (\ref{mnudia}), the complex light neutrino mass eigenvalues  are evaluated as 
\begin{eqnarray}
m^c_{1,3}&=&\left[-b \pm \sqrt{a^2-ad+d^2}\right], \\
 m^c_{2}&=&(a+d). 
\end{eqnarray}
Correspondingly the real and positive mass eigenvalues of light neutrinos are determined as 
\begin{eqnarray}
 m_1&=& \alpha \frac{y}{k}\left[\left(P-\frac{k}{y}\right)^2+Q^2 \right]^{1/2},\label{m1}\\
 m_2&=&\alpha \frac{y}{k}\left[1+\epsilon^2+2\epsilon\cos(\phi_{ab}-\phi_{db}) \right]^{1/2},\label{m2}\\
 m_3&=&\alpha \frac{y}{k}\left[\left(P+\frac{k}{y}\right)^2+Q^2 \right]^{1/2}\label{m3},
\end{eqnarray}
where 
\begin{equation}
 \alpha = \frac{k}{\Lambda}v^2\epsilon, ~~P=\left[\frac{1}{2}(A+\sqrt{A^2+B^2}) \right]^{1/2}\hspace{.05cm} {\rm and} \hspace{.08cm}
 Q=\left[\frac{1}{2}(-A+\sqrt{A^2+B^2}) \right]^{1/2},
\end{equation}
with 
\begin{eqnarray}
 A&=&(\cos 2\phi_{ab}+\epsilon^2\cos 2\phi_{db}-\epsilon\cos(\phi_{ab}+ \phi_{db})), \\
 B&=&(\sin 2\phi_{ab}+\epsilon^2\sin 2\phi_{db}-\epsilon\sin(\phi_{ab}+ \phi_{db})).
\end{eqnarray}
Also, phases ($\gamma_i$) associated with each mass eigenvalues can be expressed as 
\begin{eqnarray}
 \gamma_1&=&\phi_b+\tan^{-1}\left(\frac{Q}{P-\frac{k}{y}}\right),\\
 \gamma_2&=&\phi_b+\tan^{-1}\left(\frac{\sin\phi_{ab}+\epsilon\sin\phi_{db}}{\cos\phi_{ab}+\epsilon\cos\phi_{db}}\right),\\
 \gamma_3&=&\pi + \phi_b  +\tan^{-1}\left(\frac{Q}{P+\frac{k}{y}}\right).
\end{eqnarray}

Using the above expressions of absolute neutrino masses, we define the ratio of solar to atmospheric 
mass-squared differences as $r$, 
\begin{equation}\label{exp:r}
 r = \frac{\Delta m^2_{\odot}}{\vert \Delta m^2_{atm} \vert}, 
\end{equation}
with  $\Delta m^2_{\odot} \equiv \Delta m^2_{21} = m^2_2 - m^2_1$ and  
$\vert \Delta m^2_{atm} \vert \equiv |\Delta m^2_{31}| = |m^3_3 - m^2_1|$ .
Then it turns out that both $r$ and $\theta_{13}$ depends on $\epsilon, y/k$ and the relative 
phases: $\phi_{ab}, \phi_{db}$. The Dirac CP phase $\delta$ is also a function of these parameters 
only. As values of $r$ and $\theta_{13}$ are precisely known from neutrino oscillation data, it would 
be interesting to constrain the parameter space of  $\epsilon, y/k$ and the relative 
phases which can be useful in predicting $\delta$. However analysis with all these four parameters is 
difficult to perform. So, below we categorize few cases depending on some specific choices of 
relative phases.  In doing the analysis, following \cite{Forero:2014bxa}, the best fit values of 
$\Delta m^2_{\odot} = 7.6\times 10^{-5} $ eV$^2$ and $\vert\Delta m^2_{atm}\vert = 2.48 
\times 10^{-3} $ eV$^2$ are used for our analysis. $r$ and $\sin\theta_{13}$ are taken as 0.03 
and 0.1530 (best fit value~\cite{Forero:2014bxa}) respectively.

\subsection{Case A : $\phi_{ab}=\phi_{db}=0$}
\begin{figure}[!h]
$$
\includegraphics[height=4.5cm]{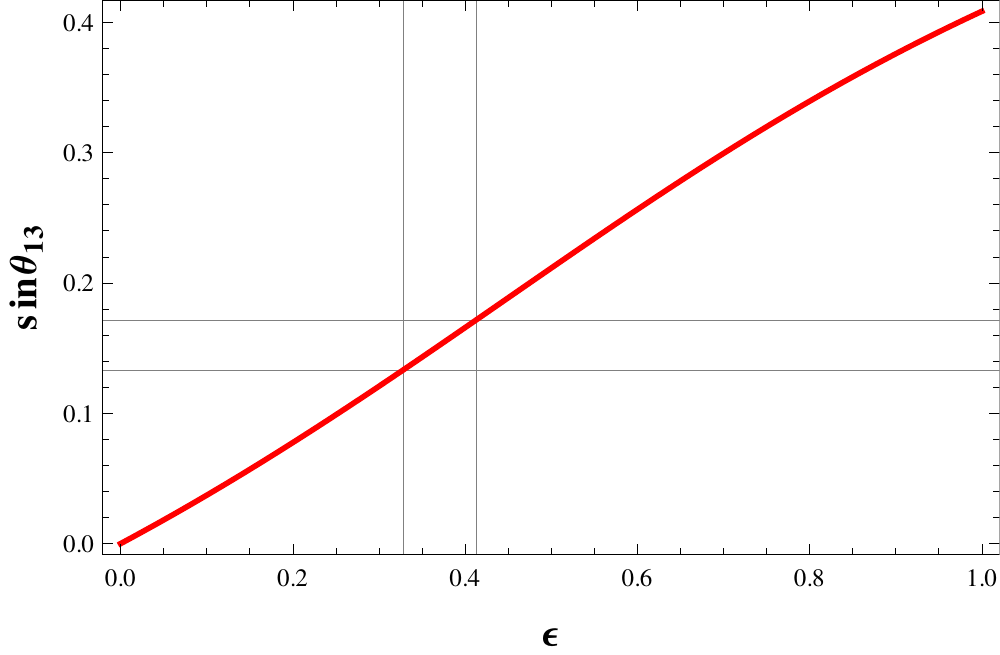}
$$
\caption{ Plot of $\sin\theta_{13}$ against $\epsilon$. 3$\sigma$ range~\cite{Forero:2014bxa} of 
$\sin\theta_{13}$ (indicated by the horizontal lines) fixes $\epsilon$ in the range: 0.328-0.4125 (indicated by vertical lines). }
\label{fig:s}
\end{figure}

Here we make the simplest choice for the phases, $\phi_{ab}=\phi_{db}=0$.  Then the Eq. (\ref{tpsi}) becomes function of $\epsilon$
alone~\cite{Karmakar:2014dva} as:
\begin{equation}
 \tan 2 \theta_{\nu}  = \frac{\sqrt{3}\epsilon}{\epsilon-2}.
\end{equation}
Hence $\sin\theta_{13}$ depends only on $\epsilon$ where following Eq. (\ref{delta-phid}), the Dirac CP phase is zero or $\pi$. 
The $\epsilon$ dependence of $\sin\theta_{13}$ is represented in Fig. \ref{fig:s}. The horizontal patch in 
Fig. \ref{fig:s} denotes the allowed 3$\sigma$ range of $\sin\theta_{13}$ ($\equiv$ 0.1330-0.1715)~\cite{Forero:2014bxa}
which is in turn restrict the range of $\epsilon$ parameter (between 0.328 and 0.4125) denoted by the 
vertical patch in the same figure. Note that the interaction strength of DM with the SM particles depends 
on $\epsilon^n\equiv Y$. Therefore we find that the size of $\sin\theta_{13}$ is intimately related with 
the Higgs portal coupling of DM. This is the most significant observation of this paper. 
With the above mentioned range of $\epsilon$, obtained from Fig. \ref{fig:s}, the two other mixing angles 
$\theta_{12}$ and $\theta_{23}$ are found to be within the 3$\sigma$ range. 

\begin{figure}[h]
$$
\includegraphics[height=4.5cm]{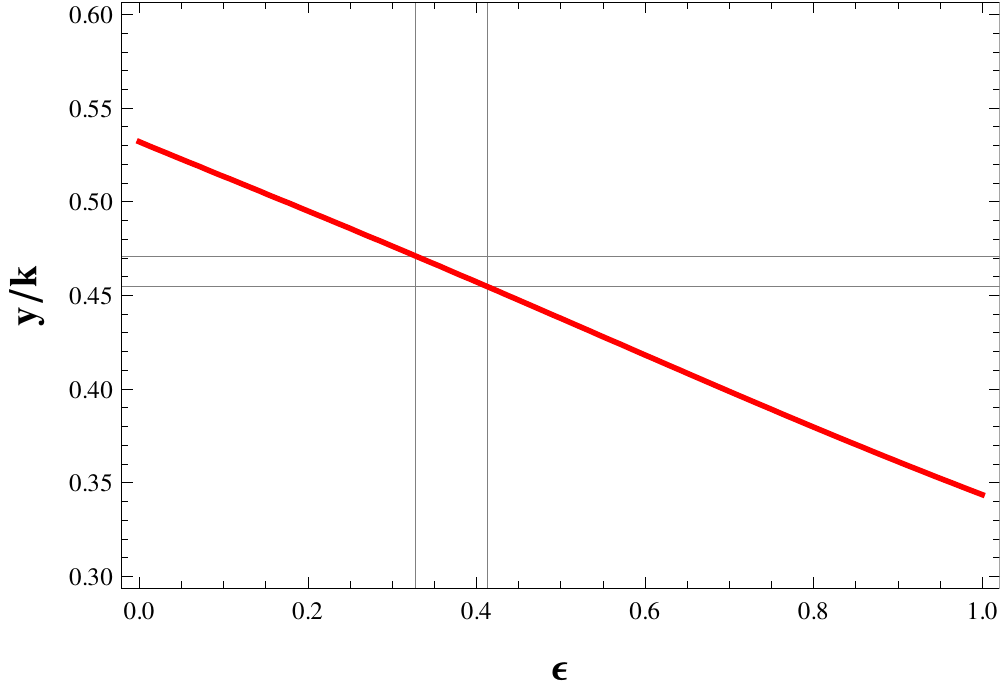}
$$
\caption{ Contour plot of $r = 0.03$ in $y/k - \epsilon$ plane. 
          The vertical lines represent the allowed range for $\epsilon$ (0.328-0.4125) corresponding to 
          3$\sigma$ range of $\sin\theta_{13}$  which restricts the ratio $y/k$ between 0.471 to 0.455 indicated 
          by horizontal lines.}
\label{fig:r0}
\end{figure}

Expressions for the real and positive mass eigenvalues are obtained from Eq. (\ref{m1}-\ref{m3})  and 
can be written as 
\begin{eqnarray}
 m_1&=&\alpha \frac{y}{k} \left|\sqrt{1-\epsilon+\epsilon^2} -k/y\right|,\label{m10}\\
 m_2&=&\alpha \frac{y}{k}\left[1+\epsilon \right],\label{m20}\\
 m_3&=&\alpha \frac{y}{k} \left[\sqrt{1-\epsilon+\epsilon^2} +k/y \right]\label{m30}. 
\end{eqnarray}
With the above mass eigenvalues, one can write the ratio of solar to atmospheric 
mass-squared differences as defined in Eq. (\ref{exp:r}) as: 
\begin{equation}
 r=\frac{3\epsilon \frac{y}{k}-\frac{k}{y}+2\sqrt{1-\epsilon+\epsilon^2}}{4\sqrt{1-\epsilon+\epsilon^2}}.
\end{equation}
From Fig. \ref{fig:s}, we have fixed $\epsilon$ range corresponding to 3$\sigma$ range of $\sin\theta_{13}$. Now, 
to satisfy $r=0.03$~\cite{Forero:2014bxa}, we vary the ratio of the coupling constants, $y/k$, against 
$\epsilon$ using Eq. (\ref{exp:r}) and (\ref{m10}-\ref{m30}). The result is presented in Fig. \ref{fig:r0}. 
The 
vertical patch there represents allowed region for $\epsilon$ fixed from Fig. \ref{fig:s} which determines the 
range of $y/k$ to be within 0.471-0.455.
After obtaining $\epsilon$ and the ratio $y/k$, we can now find the factor $k/\Lambda$ (within $\alpha$) 
in order to satisfy the solar mass-squared difference 
$\Delta m^2_{\odot}= m^2_2 - m^2_1 = 7.6\times 10^{-5} $ eV$^2$~\cite{Forero:2014bxa}. 
Using Eq. (\ref{m10}) and (\ref{m20}) we find this factor to be 
\begin{equation}
 \frac{k}{\Lambda}=\frac{1}{v^2\epsilon \frac{y}{k}} \sqrt{\frac{\Delta m^2_{\odot}}{\left[3 \epsilon - 
 \left(\frac{k}{y}\right)^2 + 2\frac{k}{y}\sqrt{1+\epsilon^2-\epsilon} \right]}}.
\end{equation}
Considering the 3$\sigma$ variation of $\sin\theta_{13}$, it falls within $1.97\times 10^{-15}$ GeV$^{-1}$ 
to $1.60\times 10^{-15}$ GeV$^{-1}$ with $v=246 $ GeV.
\begin{figure}[h]
$$
\includegraphics[height=4.5cm]{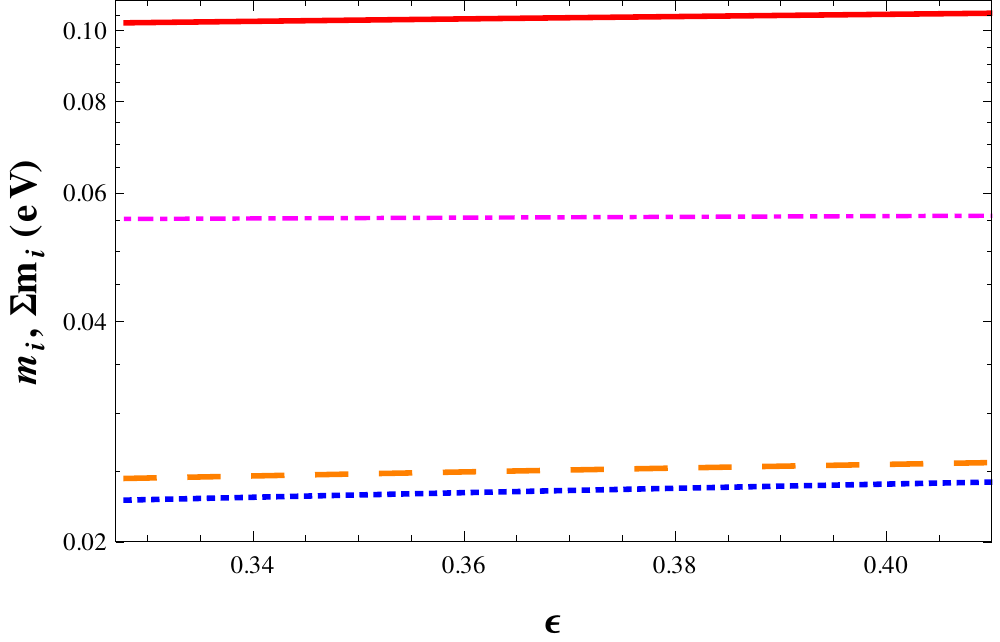}\hspace{.2cm}
\includegraphics[height=4.5cm]{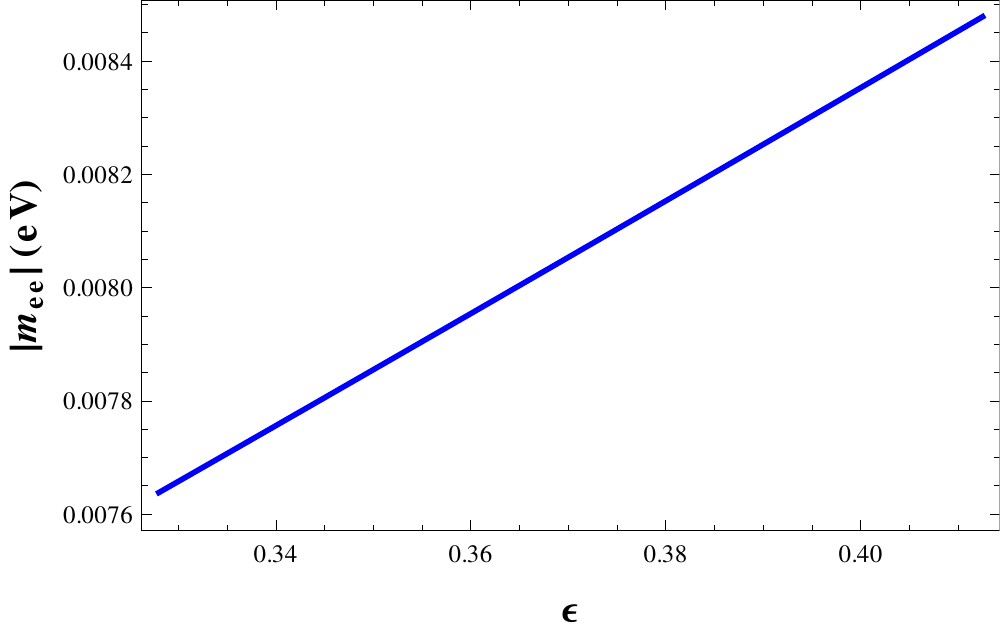}
$$
\caption{ Left: Individual absolute neutrino masses ($m_1$- blue dotted line, $m_2$- orange dashed line, 
          $m_3$- magenta dot-dashed line) and their sum (continuous red line) against $\epsilon$ (0.328-0.4125)
          corresponding to 3$\sigma$ range of $\sin\theta_{13}$. Right: Effective neutrino mass
parameter (continuous blues line)  against $\epsilon$ (0.328-0.4125) corresponding to 3$\sigma$ range of $\sin\theta_{13}$.}
\label{fig:mi0mee}
\end{figure}
\begin{table}[h]
\centering
\begin{tabular}{c|c}
\hline
Parameters/Observable & Allowed Range\\
\hline
$\epsilon$         & 0.328-0.4125  \\
$k/\Lambda$ (GeV$^{-1}$)  & $1.97\times 10^{-15}$ - $1.60\times 10^{-15}$   \\
$\Sigma m_i$ (eV)  & 0.102 - 0.106\\
$|m_{ee}|$   (eV)  & 0.00764-0.00848\\
\hline\hline
\end{tabular}\caption{\label{tab:2} {\small Range of $\epsilon, k/\Lambda, \Sigma m_i, |m_{ee}|$ for 
                                     3$\sigma$ range of $\sin\theta_{13}$ with $\phi_{ab}=\phi_{db}=0$.}}			
\end{table}
Once we know about all parameters involved like $\epsilon, y/k, k/\Lambda$ with the specific choice of the 
phases (in this case $\phi_{ab} = \phi_{db} = 0$), it is straightforward to determine absolute neutrino masses
and effective neutrino mass parameter involved in neutrinoless double beta decay using
\begin{eqnarray}
\left|m_{ee}\right|=\left|m_1^2c_{12}^2c_{13}^2+
m_2^2s_{12}^2c_{13}^2e^{i\alpha_{21}}+
m_3^2s_{13}^2e^{i(\alpha_{31}-2\delta)}\right| 
\end{eqnarray}
as shown 
in Fig. \ref{fig:mi0mee}. We also have listed the summary of the predictions of these quantities in Table \ref{tab:2}.

\subsection{Case B : $\phi_{db}=0$}
Now we consider the case: $\phi_{db}=0$. Then the relations for $\theta_{\nu}$ and $\delta$ take the form
\begin{eqnarray}
\tan{2\theta_{\nu}} 
                    &=&\frac{ 
                    \sqrt{3}\epsilon
                    }
                    {
                    (\epsilon-2\cos\phi_{ab})\cos\varphi
                    }\,,\label{thetaab}\\
\tan\delta&=&-\frac{y}{k}\sin\phi_{ab}\label{deltaab}\,.
\end{eqnarray}
So from Eqs. (\ref{ang}, \ref{thetaab}-\ref{deltaab}) and since $\tan\delta = \tan\varphi$, it is clear that unlike 
the Case A, here $\sin\theta_{13}$ depends not only on $\epsilon$ and $y/k$ but also on the phase present in the 
theory, $i.e.$ $ \phi_{ab}$. Therefore there would exist a one to one correspondence between $\epsilon$ and $y/k$ 
in order to produce a specific value of $\sin\theta_{13}$ once a particular choice of $\delta$ has been made. 

\begin{figure}[h]
$$
\includegraphics[height=5.5cm]{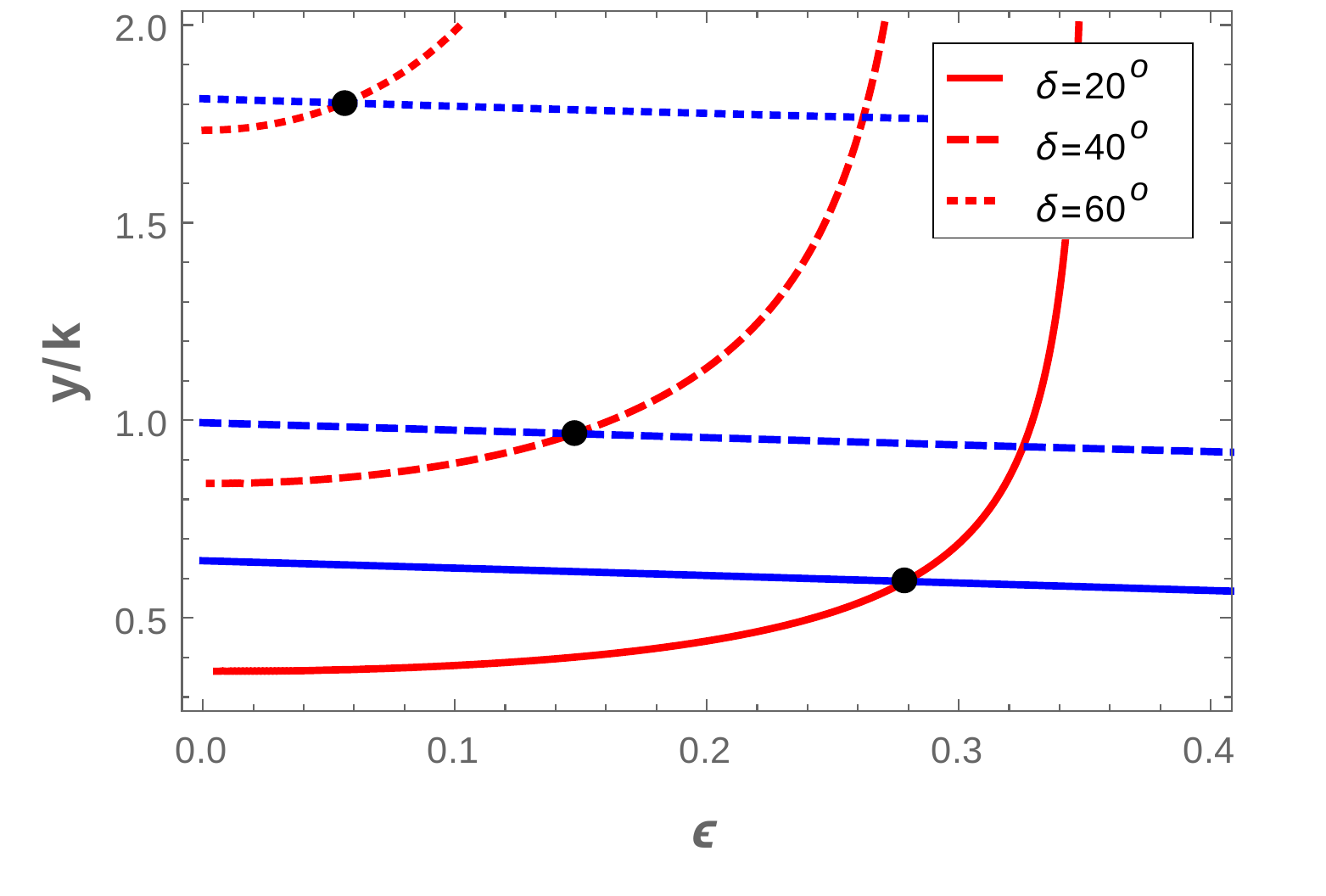}
$$
\caption{ Contour plots for both $\sin\theta_{13}=0.1530$ (shown in red continuous, dashed and dotted lines) and $r=0.03$ 
          (shown in blue continuous, dashed and dotted lines) for $\delta=20^{\circ}, \delta=40^{\circ}$ and $\delta=60^{\circ}$
          respectively in $\epsilon$-$y/k$ plane. Black dots on each intersection represents solution for $\epsilon$ and $y/k$ corresponding 
          to each $\delta$ for $\phi_{db}=0$.} 
\label{fig:db0}
\end{figure}
Now, with $\phi_{db}=0$, absolute neutrino masses given in Eq. (\ref{m1}-\ref{m3}) are reduced to 
\begin{eqnarray}
 m_1&=&\alpha \frac{y}{k} \left[ (P_1-\frac{k}{y})^2 +Q_1^2\right]^{1/2},\label{m1db0}\\
 m_2&=&\alpha \frac{y}{k} \left[1+\epsilon^2+2\epsilon\cos\phi_{ab}\right]^{1/2},\label{m2db0}\\
 m_3&=&\alpha \frac{y}{k} \left[(P_1+\frac{k}{y})^2 +Q_1^2 \right]^{1/2}\label{m3db0}, 
\end{eqnarray}
with 
\begin{equation}
 P_1=\left[\frac{1}{2}(A_1+\sqrt{A_1^2+B_1^2}) \right]^{1/2}\hspace{.05cm} , \hspace{.08cm}
 Q_1=\left[\frac{1}{2}(-A_1+\sqrt{A_1^2+B_1^2}) \right]^{1/2}, 
\end{equation}
\begin{equation}
 A_1=\left(\epsilon^2+\cos 2\phi_{ab}-\epsilon\cos\phi_{ab} \right)\hspace{.05cm} {\rm and} \hspace{.08cm}
 B_1=\left(\sin 2\phi_{ab}-\epsilon\sin\phi_{ab} \right).  
\end{equation}
The ratio of solar to atmospheric neutrino mass-squared differences takes the form
\begin{equation}
 r=\frac{1}{4P_1\frac{k}{y}}\left[(1+\epsilon^2+2\epsilon\cos\phi_{ab}) -\left(P_1-\frac{k}{y}\right)^2 -Q_1^2\right].
\end{equation}
Clearly, one finds that $\epsilon$ and $y/k$ are the only parameters involved in both  $\sin\theta_{13}$ and $r$ 
once $\delta$ values are taken as input. Therefore, those values of $\epsilon$ and $y/k$ are allowed which simultaneously
satisfy data obtained for $\sin\theta_{13}$ and $r$ from neutrino oscillation experiments. Here we have considered the 
best fit values from~\cite{Forero:2014bxa} and drawn contour plots for $\sin\theta_{13}=0.1530$ and
$r=0.03$. Intersection of these contours then represents solutions for $\epsilon$ and $y/k$. Note that 
$\delta = 0$ case corresponds to the results obtained in Case A. 

In Fig. \ref{fig:db0}, we have plotted typical contours obtained for $\sin\theta_{13}=0.1530$ (red lines) and $r=0.03$ (blue lines) for 
$\delta=20^{\circ}, \delta=40^{\circ}$ and $\delta=60^{\circ}$ respectively in $\epsilon$-$y/k$ plane. The intersecting points are denoted 
by black dots and represent the solution points for $\epsilon$ and $y/k$. In Table {\ref{tab:3}} we have listed estimations for 
$\epsilon$ and $y/k$ for different $\delta$ values. 
\begin{table}[h]
\centering
\begin{tabular}{cccccccccc}
\hline
$\delta$ & $\epsilon$ & $y/k$  &  $k/\Lambda$ ($10^{-15}$ GeV$^{-1}$) & $\Sigma m_i$ (eV) &$|m_{ee}|$ (eV)\\
\hline
$0^{\circ}$  & 0.372 & 0.463 & 1.756 & 0.1042& 0.0222 \\
$10^{\circ}$ & 0.343 & 0.496 & 1.910 & 0.1068& 0.0236 \\
$20^{\circ}$ & 0.279 & 0.592 & 2.361 & 0.1143& 0.0274 \\
$30^{\circ}$ & 0.209 & 0.745 & 3.140 & 0.1267& 0.0331 \\
$40^{\circ}$ & 0.147 & 0.966 & 4.405 & 0.1454& 0.0409 \\
$50^{\circ}$ & 0.096 & 1.288 & 6.610 & 0.1743& 0.0516 \\
$60^{\circ}$ & 0.056 & 1.803 & 11.10 & 0.2230& 0.0682 \\
$61^{\circ}$ & 0.053 & 1.873 & 11.80 & 0.2298& 0.0704 \\
\hline
$70^{\circ}$ & 0.026 & 2.798 & 23.22 & 0.3210& 0.1002 \\
$80^{\circ}$ & 0.007 & 5.743 & 85.42 & 0.6173& 0.1952 \\
\hline\hline
\end{tabular}\caption{\label{tab:3} {\small Estimated values of various parameters and observables satisfying neutrino 
oscillation data for different values of $\delta$ with $\phi_{db}=0$ .}}			
\end{table}
Just like the previous case, after obtaining $\epsilon$ and $y/k$, we can find the factor $k/\Lambda$ using the fact 
that it has to produce correct solar mass-squared difference $\Delta m^2_{\odot}= m^2_2 - m^2_1 = 7.6\times 10^{-5} $ 
eV$^2$~\cite{Forero:2014bxa}. For this, we employ Eq. (\ref{m1db0}) and (\ref{m2db0}).  All these findings are mentioned in Table \ref{tab:3} including sum of the 
absolute masses ($\Sigma m_i$) of all three light neutrinos and effective neutrino mass parameter involved in  neutrinoless
double beta decay ($|m_{ee}|$) for different considerations of leptonic CP phase $\delta$. In this analysis we observe that,
for various values of $\delta$ between $0^{\circ}$ to $360^{\circ}$ there are certain points where same set of solutions
for $\epsilon$ and $y/k$ are repeated ($\it e.g.$ solutions with $\delta$ is repeated for $|\pi-\delta|$). 
We should also employ the upper bound of sum of all three 
light neutrino masses ($\Sigma m_i<0.23$ eV) coming from cosmological observation by Planck~\cite{planck}. 
Once this is included, we note that some of the $\delta$ values need to be discarded as the corresponding sum of the masses 
exceeds 0.23 eV
as seen from Table 
\ref{tab:3}. We therefore conclude that the allowed values for $\delta$ are: 
between $0^{\circ}-61^{\circ}$ (and also $119^{\circ}-180^{\circ}$, $180^{\circ}-241^{\circ}$ and $299^{\circ}-360^{\circ}$). 

\subsection{Case C : $\phi_{ab}=0$}
When $\phi_{ab}=0$, relations for $\theta_{\nu}$ and $\delta$ take the form
\begin{eqnarray}
\tan{2\theta_{\nu}} 
                    &=&\frac{ 
                    \sqrt{3}\epsilon\cos\phi_{db}
                    }
                    {
                    (\epsilon\cos\phi_{db}-2)\cos\varphi
                    },\label{thetadb}\\
\tan\delta&=&\frac{y}{k}\tan\phi_{ab}\label{deltadb}.
\end{eqnarray}
Here also $\sin\theta_{13}$ depends on $\epsilon, y/k$ and the phase involved $\phi_{db}$.
The real and positive mass eigenvalues can be written as
\begin{eqnarray}
 m_1&=&\alpha \frac{y}{k}  \left[ (P_2-\frac{k}{y})^2 
+Q_2^2\right]^{1/2},\label{m1ab0}\\
 m_2&=&\alpha \frac{y}{k} 
\left[1+\epsilon^2+2\epsilon\cos\phi_{db}\right]^{1/2},\label{m2ab0}\\
 m_3&=&\alpha \frac{y}{k}  \left[(P_2+\frac{k}{y})^2 +Q_2^2 
\right]^{1/2}\label{m3ab0}, 
\end{eqnarray}
with 
\begin{equation}
 P_2=\left[\frac{1}{2}(A_2+\sqrt{A_2^2+B_2^2}) \right]^{1/2}\hspace{.05cm} , 
\hspace{.08cm}
 Q_2=\left[\frac{1}{2}(-A_2+\sqrt{A_2^2+B_2^2}) \right]^{1/2},
\end{equation}
where 
\begin{equation}
 A_2=\left(1+\epsilon^2 \cos 2\phi_{db}-\epsilon\cos\phi_{db} 
\right)\hspace{.05cm} {\rm and} \hspace{.08cm}
 B_2=\left(\epsilon^2\sin 2\phi_{db}-\epsilon\sin\phi_{db} \right).  
\end{equation}
The ratio of solar to atmospheric neutrino mass-squared differences takes the form
\begin{equation}
 r=\frac{y/k}{4P_2}\left[(1+\epsilon^2+2\epsilon\cos\phi_{db}) -(P_2-k/y)^2 
-Q_2^2\right].
\end{equation}
\begin{figure}[h]
$$
\includegraphics[height=5.5cm]{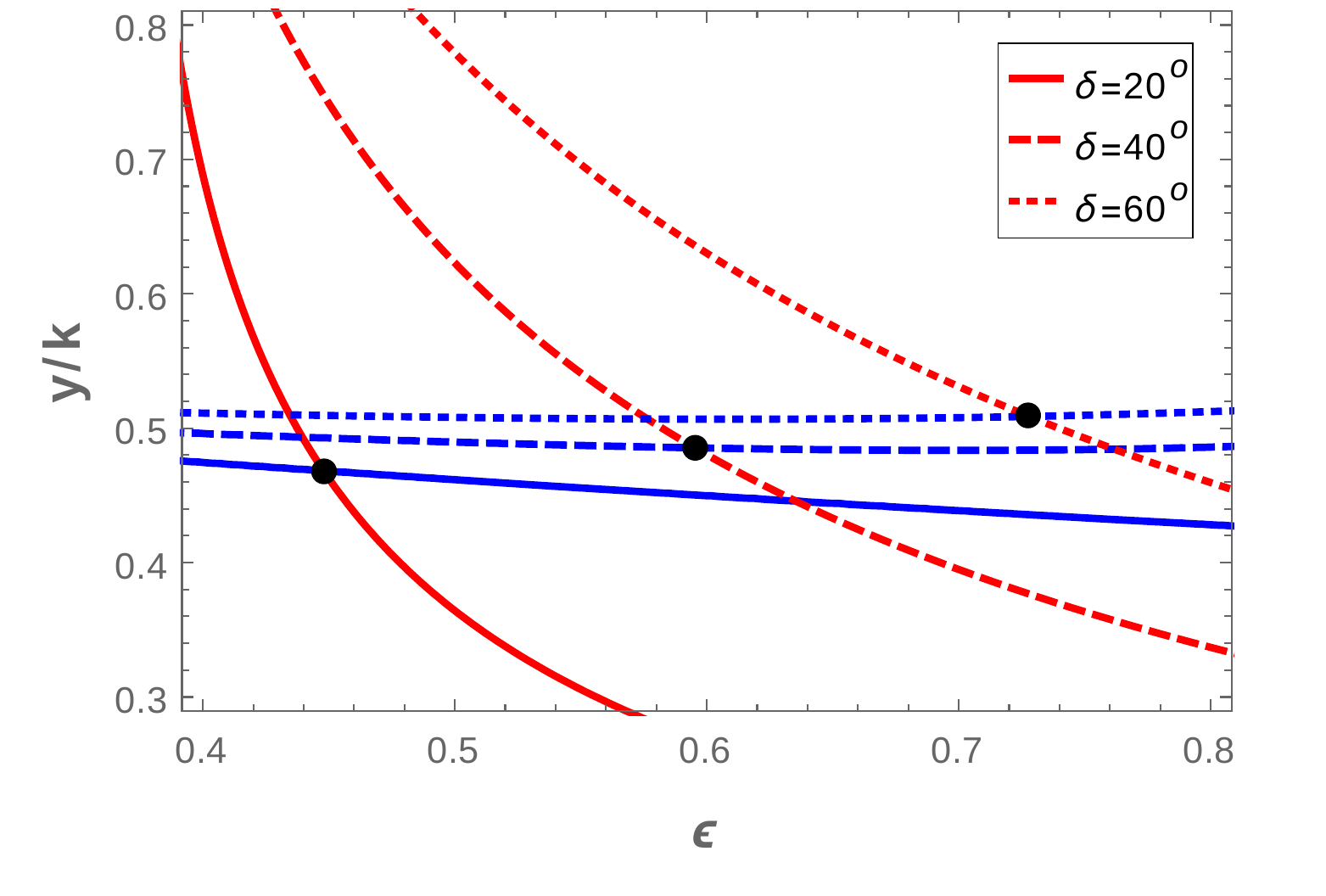}
$$
\caption{ Contour plots for both $\sin\theta_{13}=0.1530$ (shown in red continuous, dashed and dotted lines) and $r=0.03$ 
          (shown in blue continuous, dashed and dotted lines) for $\delta=20^{\circ}, \delta=40^{\circ}$ and $\delta=60^{\circ}$
          respectively in $\epsilon$-$y/k$ plane. Black dots on each intersection represents solution for $\epsilon$ and $y/k$ corresponding 
          to each $\delta$ for $\phi_{ab}=0$.}
\label{fig:ab0}
\end{figure}

We then scan the parameter space for $\epsilon$ and $y/k$ for various choices of $\delta$ so as to have $r = 0.03$ and 
$\sin\theta_{13} = 0.153$. In Fig. \ref{fig:ab0}, we 
provide contour plots for $\sin\theta_{13} = 0.1530$ (red lines) and $r = 0.03$ (blue lines) for 
$\delta=20^{\circ}, \delta=40^{\circ}$ and $\delta=60^{\circ}$. The intersection between $\sin\theta_{13}$ and $r$ contours indicate the 
simultaneous satisfaction of them. Hence the intersections are indicated by black dots with which a pair of $\epsilon, y/k$ 
are attached. Similar to the previous two cases, here we estimate the $k/\Lambda$ for each such pair of $\epsilon, y/k$ 
with a specific $\delta$. This in turn provide an estimate of $\Sigma m_i$ and effective mass parameter $|m_{ee}|$ depending on the choice of $\delta$. 
We provide these outcomes in Table \ref{tab:4}. 
\begin{table}[h]
\centering
\begin{tabular}{cccccccccc}
\hline
$\delta$ & $\epsilon$ & $y/k$  &  $k/\Lambda$ ($10^{-15}$ GeV$^{-1}$) & $\Sigma m_i$ (eV) &$|m_{ee}|$ (eV)\\
\hline
$0^{\circ}$  & 0.372 & 0.463 & 1.756 & 0.1042& 0.0222 \\
$10^{\circ}$ & 0.393 & 0.464 & 1.670 & 0.1048& 0.0225 \\
$20^{\circ}$ & 0.448 & 0.468 & 1.480 & 0.1065& 0.0233 \\
$30^{\circ}$ & 0.520 & 0.475 & 1.300 & 0.1093& 0.0245 \\
$40^{\circ}$ & 0.595 & 0.485 & 1.167 & 0.1128& 0.0260 \\
$50^{\circ}$ & 0.666 & 0.497 & 1.065 & 0.1162& 0.0273 \\
$60^{\circ}$ & 0.728 & 0.509 & 0.981 & 0.1182& 0.0280 \\
$70^{\circ}$ & 0.782 & 0.519 & 0.901 & 0.1179& 0.0275 \\
$80^{\circ}$ & 0.827 & 0.526 & 0.826 & 0.1152& 0.0259 \\
\hline\hline
\end{tabular}\caption{\label{tab:4} {\small Estimated values of various parameters and observables satisfying neutrino 
oscillation data for different values of $\delta$ with $\phi_{ab}=0$ .}}			
\end{table}

\subsection{Case D : $\phi_{ab}=\phi_{db}=\beta$}
With $\phi_{ab}=\phi_{db}=\beta$, the mixing angle
$\theta_{\nu}$ turns out to be function of $\epsilon$ only  and  is given by
\begin{equation}
 \tan 2 \theta_{\nu}  =  \frac{\sqrt{3}\epsilon}{\epsilon-2},
\end{equation}
while $\tan\delta$ becomes zero. Note that the expressions for the 
mixing angle $\theta_{\nu}$ and $\delta$ are identical to the 
ones obtained in Case A. Therefore we use the constraint on $\epsilon$ 
obtained from Fig. \ref{fig:s}
in order to satisfy $3\sigma$ allowed range of $\sin\theta_{13}$. However the expressions for 
real and positive mass eigenvalues involve the common phase $\beta$ and  can be written as (following 
Eqs. (\ref{m1}-\ref{m3}))
\begin{eqnarray}
 m_1&=&\alpha\frac{y}{k} 
\left[\left(\sqrt{1-\epsilon+\epsilon^2}\cos\beta-\frac{k}{y}\right)^2 
+\left(\sqrt{1-\epsilon+\epsilon^2}\sin\beta\right)^2\right]^{1/2},\label{dm10}\\
 m_2&=&\alpha \frac{y}{k}\left[1+\epsilon 
\right],\label{dm20}\\
 m_3&=&\alpha \frac{y}{k} 
\left[\left(\sqrt{1-\epsilon+\epsilon^2}\cos\beta+\frac{k}{y}\right)^2 
+\left(\sqrt{1-\epsilon+\epsilon^2}\sin\beta\right)^2\right]^{1/2}\label{dm30}. 
\end{eqnarray}
\begin{figure}[h]
$$
\includegraphics[height=5.0cm]{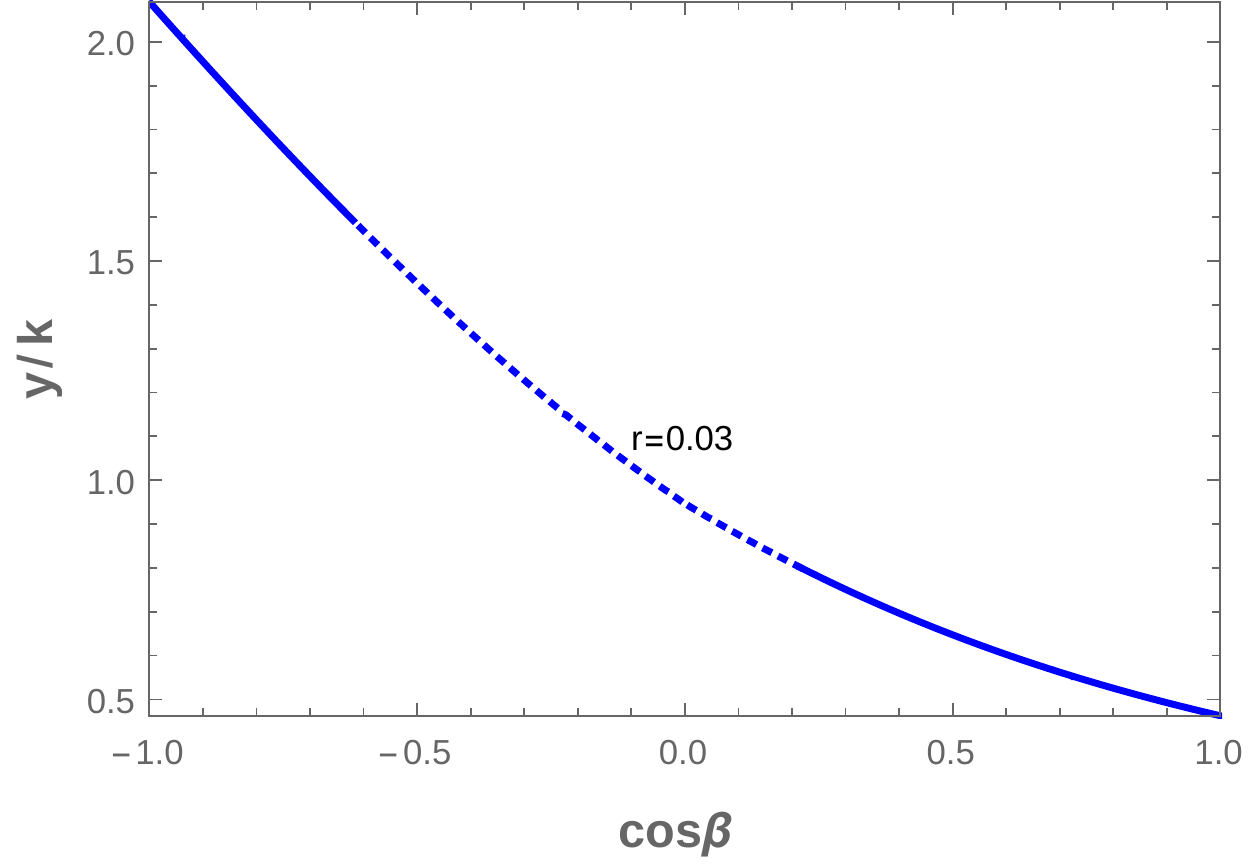}
$$
\caption{  Contour plot for $r=0.03$ in the $y/k-\cos\beta$ plane for 
           $\phi_{db}=\phi_{ab}=\beta$. The disallowed range of 
$y/k,\cos\beta$  is 
indicated by the dotted portion. }
\label{fig:cosYk}
\end{figure}
Then following our approach for finding the range of parameters which would satisfy 
the oscillation parameters obtained from experimental data, we define the ratio of solar to atmospheric 
mass-squared differences as defined in Eq. (\ref{exp:r}) as 
\begin{equation}\label{rphi}
 r=\frac{3\epsilon 
\frac{y}{k}-\frac{k}{y}+2\cos\beta\sqrt{1-\epsilon+\epsilon^2}}{4|\cos\beta|\sqrt{
1-\epsilon+\epsilon^2}}.
\end{equation}

\begin{figure}[h!]
$$
\includegraphics[height=5.0cm]{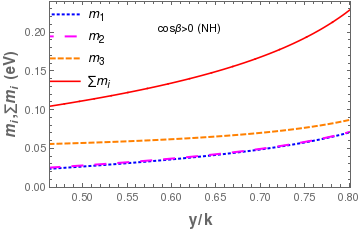}
\includegraphics[height=5.0cm]{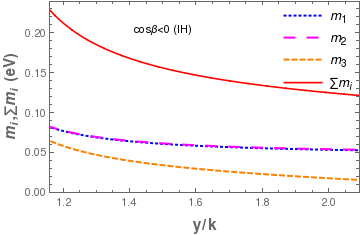}
$$
\caption{ Absolute neutrino masses vs $y/k$ (blue dotted,  magenta large-dashed,
orange dashed  and red continuous lines represent $m_1$, $m_2$, $m_3$ 
and $\sum m_{i}$ respectively). The left panel is for $\cos\beta>0$ and  
right panel is for $\cos\beta<0$.}
\label{fig:misphi}
\end{figure} 

From Fig. \ref{fig:s} we fix $\epsilon=0.372$ which would produce the best fit value of 
$\sin\theta_{13}$. Then, using the ratio of solar to atmospheric mass squared 
difference as given in Eq. (\ref{rphi}), we can constrain $y/k$ and $\cos\beta$. 
Here we plot $r=0.03$ contour in the $y/k$-$\cos\beta$ plane  as shown in Fig. 
\ref{fig:cosYk}. For $-1\leq\cos\beta\leq 1$. We observe that 
$y/k$ falls within the range: $0.463\leq\ y/k \leq 2.091$. Thus Fig. \ref{fig:cosYk} establishes a 
correlation between $y/k$ and $\cos\beta$. 
Now to find absolute neutrino masses we need to obtain 
$k/\Lambda$ first.  We can find $k/\Lambda$ from the best fit value for solar mass squared 
difference,  $m^2_2 - m^2_1 = 7.6\times 10^{-5}$ eV$^2$, and is given by 
\begin{equation}
 \left(\frac{k}{\Lambda}\right)^2=\frac{\Delta m^2_{\odot} }{4r(v^2\epsilon)^2 
y/k |\cos\beta| \sqrt{1+\epsilon^2-\epsilon}}. 
\label{kLD}
\end{equation}
We have used Eq. (\ref{dm10}-\ref{dm30}) to obtain the above equation. 
Once $\epsilon$ is fixed at 0.372 and following Fig. \ref{fig:cosYk} we know 
$y/k$ and corresponding $\cos\beta$ (to have $r = 0.03$), we can use Eq. (\ref{kLD}) 
to have an estimate for $k/\Lambda$.
Now by knowing $k/\Lambda$, we have plotted absolute 
masses for light neutrinos in Fig. \ref{fig:misphi} by using  Eq. (\ref{dm10}-\ref{dm30}). 
Here the left (right) panel is for $\cos\beta>0(<0)$ and indicates normal (inverted) hierarchy for 
light neutrino masses. In Fig. \ref{fig:misphi}, absolute neutrino masses $m_1, 
m_2, m_3$ and $\sum m_i$ are denoted by blue dotted, magenta large-dashed, 
orange dashed and red continuous lines respectively. Note that here we have 
plotted sum of the three absolute light neutrino masses consistent with the 
recent observation made by PLANCK, $i.e.$ 
$\sum m_i \leq 0.23$ eV~\cite{planck}. If we impose this 
constraint on the sum of absolute masses of the three light neutrinos, 
then the allowed region for $y/k$ gets further constrained. The dotted portion 
in Fig. \ref{fig:cosYk} represents this excluded part. Therefore the allowed 
region for $y/k$ then turns out to be $0.463\leqslant y/k \leqslant 0.802$ 
for $\cos\beta>0$ (normal hierarchy) and $1.159 \leqslant y/k \leqslant 
2.091$ for $\cos\beta<0$ (inverted hierarchy). Finally in this case, the 
prediction for $|m_{ee}|$ found to be within 
$0.022 \hspace{.1cm}{\rm eV} <|m_{ee}|<0.039 \hspace{.1cm}{\rm eV}$ for normal 
hierarchy
and $0.016 \hspace{.1cm}{\rm eV} <|m_{ee}|<0.035 \hspace{.1cm}{\rm eV}$ for 
inverted hierarchy. 
\section{Phenomenology of DM Sector}\label{DM_section}
The dark sector consists of two vector-like fermions: a fermion doublet $\psi$ and a singlet $\chi$. The 
corresponding Lagrangian respecting the $U(1)$ and other discrete symmetries is
provided in Eq. (\ref{eq:Lag-lepton}). At this stage we can remind 
ourselves about the minimality of the construction in terms of choice of constituents of the dark sector. Note that 
a vector-like singlet fermion alone can not have a coupling with the SM sector 
at the renormalizable level and thereby 
its relic density is expected to be over abundant (originated from interaction suppressed by the new physics scale 
$\Lambda$). 
On the contrary, a vector-like fermion doublet alone can have significant annihilation cross section from its gauge 
interaction with the SM sector and thereby we would expect the corresponding
dark matter relic density to be under-abundant unless the DM mass is 
exorbitantly high. 
Hence we can naturally ask the question whether involvement of a singlet and a doublet vector-like fermions can 
lead to the dark matter relic density at an acceptable level. It then crucially depends on the mixing term between the 
singlet and the doublet fermions, $i.e.$ on  $m_D = Yv$. We expect a rich phenomenology out of it particularly 
because the coupling $Y$ depends on the parameter $\epsilon$ through $Y = \epsilon^n$ where $\epsilon$ plays 
an important role in the neutrino physics as evident from our discussion in the previous section. We aim to restrict 
$n$ phenomenologically. 

The electroweak phase transition along with the $U(1)$ breaking give rise to the following mass matrix 
in the basis $(\chi^0, \psi^0)$ 
\begin{equation}
\mathcal{M} = \begin{pmatrix}  M_\chi &m_D\cr \\
m_D & M_\psi \,
\end{pmatrix}.
\end{equation} 
We obtain mass eigenstates $\psi_1$ and $\psi_2$ with masses $M_1$ and $M_2$ respectively after diagonalization 
of the above matrix as 
\begin{eqnarray}
\psi_1= \cos \theta_d \chi^0 + \sin \theta_d \psi^0,\nonumber\\
\psi_2=\cos \theta_d \psi^0 - \sin \theta_d \chi^0\,,
\label{psi-12}
\end{eqnarray}
where $\tan 2\theta_d = 2m_D/{\left( M_\psi  - M_\chi \right)}$. We will work in the regime where $m_D << 
M_\psi, M_\chi$. This choice would be argued soon. However this is not unnatural as the dark matter is 
expected to interact weakly. In this limit, the mass eigenvalues are found to be
\begin{eqnarray}
M_1 \approx M_\chi-\frac{m_D^2}{M_\psi-M_\chi}, \nonumber\\
M_2 \approx M_\psi + \frac{m_D^2}{M_\psi-M_\chi}. \label{mass-eigenstates}
\end{eqnarray}
In this small mixing limit, we can write $M_\psi-M_\chi \simeq M_2-M_1 = \Delta M$. Therefore the mixing angle 
$\theta_d$ can be approximately represented by 
\begin{equation}\label{theta-d}
\sin 2\theta_d \simeq \frac{2 Y v}{\Delta M}\,.
\end{equation}
Then as evident from Eqs. (\ref{psi-12}), $\psi_1$ is dominantly the singlet having a small admixture with the 
doublet. We assume it to be the lightest between the two ($i.e. ~M_1 < M_2$) and forms the DM component 
of the universe. In the physical spectrum, we also have a charged fermion $\psi^+ (\psi^-)$ with mass $M^+ (M^-) 
= M_1 \sin^2 \theta_d + M_2 \cos^2 \theta_d$. In the limit $ \theta_d \to 0$, $M^\pm = M_2= M_\psi$. In this section, 
we will discuss the relic density of dark matter as a function of $Y$. Although $Y$ represents Yukawa coupling of the 
DM with SM Higgs, in presence of a singlet and doublet fermions, $Y$ is also a function of the mixing angle $\sin\theta_d$
as well as the mass splitting ($\Delta M$ as in Eq. (\ref{theta-d})) which 
crucially controls DM phenomenology as we demonstrate in the following 
discussion. 

\begin{figure}[h]
\centering
\includegraphics[height=2.0cm]{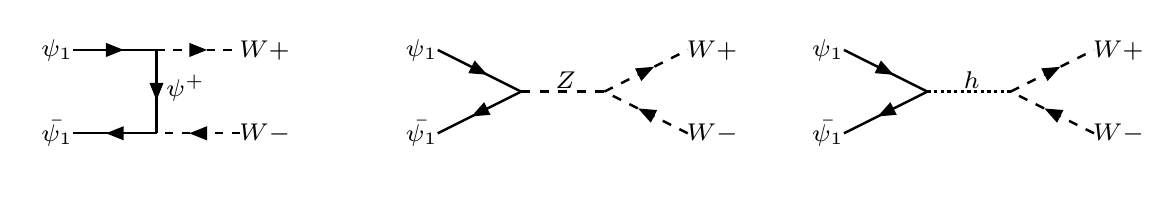}
\includegraphics[height=2.0cm]{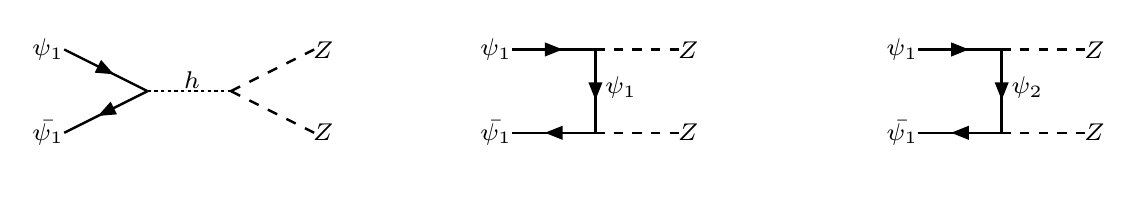}
\includegraphics[height=2.0cm]{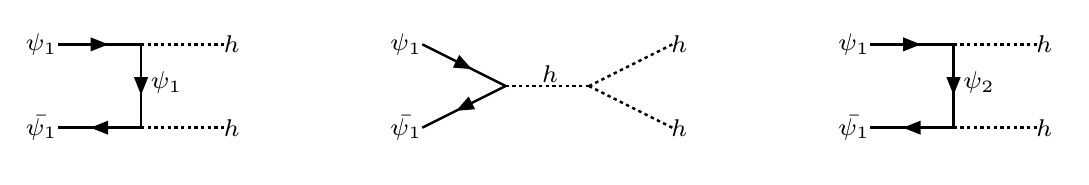}
\caption{Dominant Annihilation processes to Higgs and Gauge boson final states.}
\label{fig:FD1}
\end{figure}

Note that $\psi_0$ being the gauge doublet, it carries the gauge interactions and hence, the physical mass eigenstates including the DM have the following interaction with $Z, W$ bosons as :
\begin{equation}
\frac{g}{\sqrt{2}} \overline{\psi_0} \gamma^\mu W_\mu^+ \psi^- + {\rm h.c.} \to \frac{g\sin \theta_d}{\sqrt{2}} 
\overline{\psi_1}\gamma^\mu W_\mu^+ \psi^- + \frac{g\cos \theta_d}{\sqrt{2}} \overline{\psi_2}\gamma^\mu W_\mu^+ \psi^- + {\rm h.c.}\,,
\end{equation} 
\begin{align}
\frac{g}{2\cos \theta_w} \overline{\psi_0}\gamma^\mu Z_\mu \psi_0  & \to \frac{g}{2\cos \theta_w} \left( \sin^2 \theta_d \overline{\psi_1} \gamma^\mu Z_\mu \psi_1 
+ \sin \theta_d \cos \theta_d ( \overline{\psi_1} \gamma^\mu Z_\mu \psi_2 + \overline{\psi_2} \gamma^\mu Z_\mu \psi_1)
\right. 
\notag\\
   &~~~~~~~~~~~~~~~~~~~~~~~~~~~~~~~~~~~~~~~~~~~~ \left.
   + \cos^2 \theta_d \overline{\psi_2} \gamma^\mu Z_\mu \psi_2 \right) \,.
\end{align}

\begin{figure}[thb]
\centering
\includegraphics[height=2.0cm]{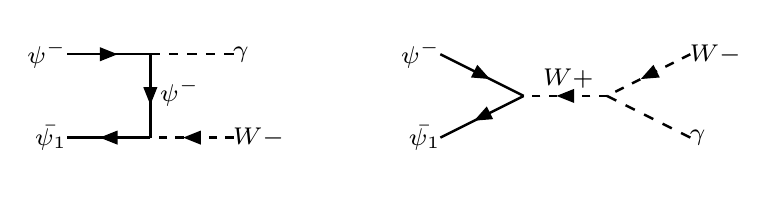}
\includegraphics[height=2.0cm]{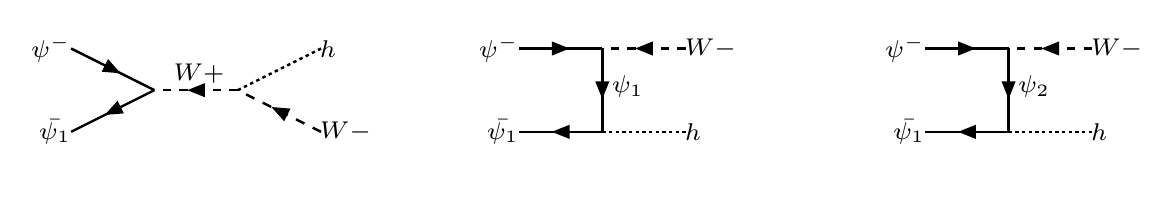}
\includegraphics[height=2.0cm]{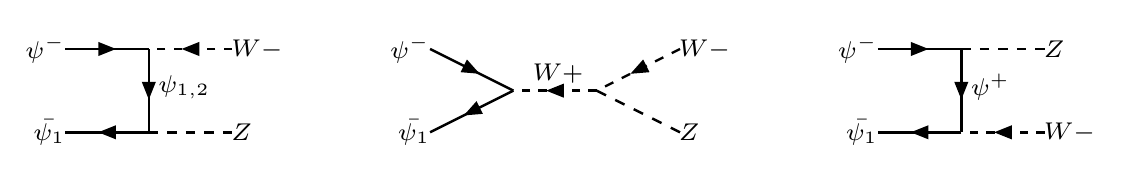}
\caption{Dominant Co-Annihilation $\bar{\psi_1}\psi^-$ to Higgs and gauge boson final states.}
\label{fig:FD3}
\end{figure}

\begin{figure}[thb]
\centering
\includegraphics[height=1.7cm]{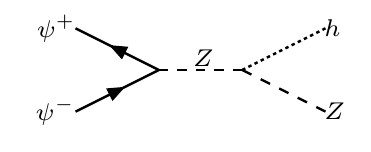}
\includegraphics[height=1.7cm]{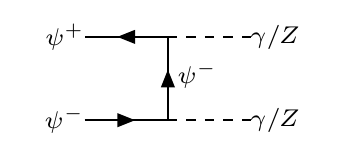}
\includegraphics[height=1.7cm]{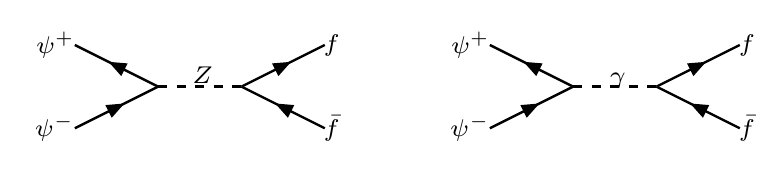}
\includegraphics[height=1.7cm]{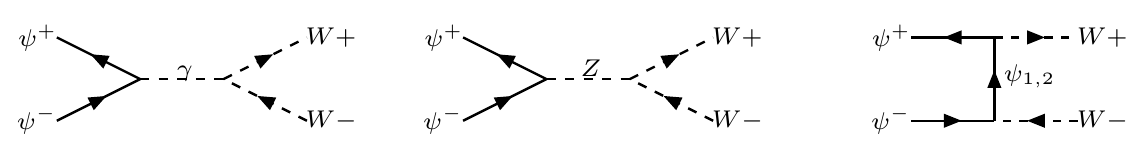}
\caption{Dominant Co-Annihilation processes of ${\psi^+}\psi^- \rightarrow$  SM particles where $f$ represents 
SM fermions.}
\label{fig:FD4}
\end{figure}

The relic density of the dark matter ($\psi_1$) is mainly dictated by annihilations through (i) $\overline{\psi_1}\psi_1 \to W^+W^-, ZZ$ through $SU(2)_L$ gauge coupling and (ii) 
$\overline{\psi_1} \psi_1 \to h h$ through Yukawa coupling introduced in Eq. (\ref{lagrangian}). The relevant processes 
\begin{figure}[thb]
\centering
\includegraphics[height=2.0cm]{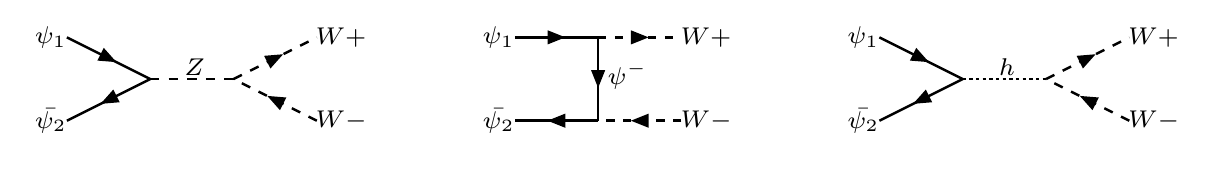}
\includegraphics[height=2.0cm]{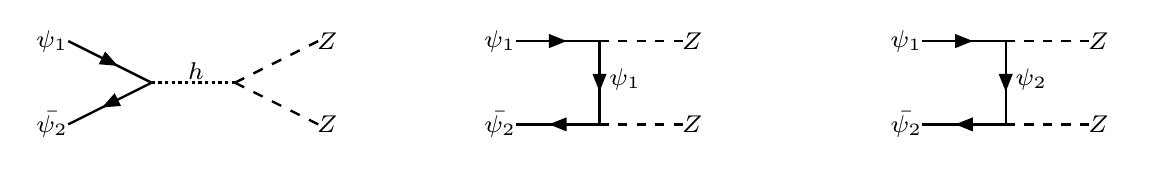}
\includegraphics[height=2.0cm]{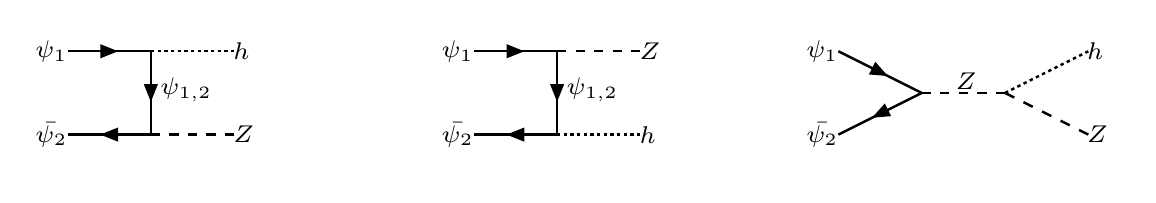}
\includegraphics[height=2.0cm]{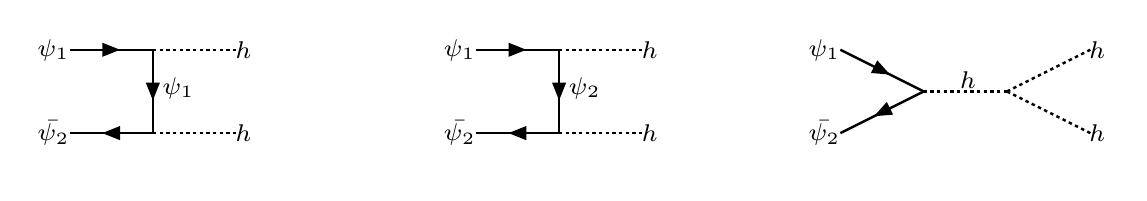}
\caption{Dominant Co-Annihilations $\psi_1\bar{\psi_2}$ to Higgs and Gauge boson final states.}
\label{fig:FD2}
\end{figure}
are indicated in Fig. \ref{fig:FD1}.  The other possible channels are mainly 
co-annihilation of $\psi_1$ with $\psi_2$  
(see Fig. \ref{fig:FD2}),  $\psi_1$ with $\psi^\pm$  (see Fig. \ref{fig:FD3}) 
and annihilations of $\psi^\pm$ 
(see Fig. \ref{fig:FD4}) which would dominantly contribute to relic density 
in a large region of parameter space~\cite{Bhattacharya:2015qpa,griest,Cynolter:2008ea, Cohen:2011ec,Cheung:2013dua} 
as can be seen once we proceed further. At this stage we can argue on our choice of making $\theta_d$ small, or in other words why 
the mixing with doublet is necessary to be small for the model to provide a DM with viable relic density. This is because the larger is 
the doublet content in DM $\psi_1$, the annihilation goes up significantly in particular through $\psi_1 \overline{\psi_1} 
\to W^{+}W^{-}$ through $Z$ and hence yielding a very small relic density. So in the small mixing limit, $\psi_2$ is dominantly a 
doublet having a mixture of minor singlet component. This implies that $\psi_2$ mass is required to be larger than 
45 GeV in order not to be in conflict with the invisible $Z$-boson decay width.

The relic density of the $\psi_1$ DM with mass $M_1$ can be given by~\cite{griest}
\begin{equation}\label{eq:omega}
\Omega_{\psi_1} h^2 = \frac{1.09 \times 10^9 \rm~ GeV^{-1}}{g_\star ^{1/2} M_{PL}} \frac{1}{J(x_f)}\,,
\end{equation}
where $J(x_f)$ is given by
\begin{equation}
J(x_f)= \int_{x_f}^ \infty \frac{\langle \sigma |v| \rangle _{eff}}{x^2} \hspace{.2cm} dx \,.
\end{equation}
Here $\langle \sigma |v| \rangle _{eff}$ is the thermal average of dark matter annihilation cross sections including 
contributions from co-annihilations as follows\footnote{If $M_2$ is very close 
to $M_1$ then $\psi_2$ decay to $\psi_1$ should contribute to relic density. 
However the parameter space scan that we have performed with $\Delta M \gtrsim 
$ GeV, excludes such possibility.}:
\begin{equation}\label{sigma_eff}
\begin{split}
\langle \sigma |v| \rangle _{eff} = & \frac{g_1^2}{g_{eff}^2} \sigma (\overline{\psi_1} \psi_1)
 +2 \frac{g_1 g_2}{g_{eff}^2} \sigma (\overline{\psi_1} \psi_2) (1+\Delta)^{3/2} exp(-x\Delta) \\
 & +2 \frac{g_1 g_3}{g_{eff}^2} \sigma (\overline{\psi_1} \psi^-) (1+\Delta)^{3/2} exp(-x\Delta)\\
 &+2 \frac{g_2 g_3}{g_{eff}^2} \sigma (\overline{\psi_2} \psi^-) (1+\Delta)^{3} exp(-2x\Delta)
 + \frac{g_2 g_2}{g_{eff}^2} \sigma (\overline{\psi_2} \psi_2) (1+\Delta)^{3} exp(-2x\Delta) \\
& + \frac{g_3 g_3}{g_{eff}^2} \sigma (\psi^+ \psi^-) (1+\Delta)^{3} exp(-2x\Delta)\,.
\end{split}
\end{equation}
In the above equation $g_1$,$g_2$ and $g_3$ are the spin degrees of freedom for $\psi_1$, $\psi_2$ and $\psi^-$ respectively. Since these are spin half 
particles, all $g$'s are 2. The freeze-out of $\psi_1$ is parameterised by $x_f= \frac{M_{1}}{T_f}$, where $T_f$ is the freeze out temperature. 
$\Delta$ depicts the mass splitting ratio as $\Delta = \frac{M_{2}- M_{1}}{ M_{1}} = \frac{\Delta M}{M_1} $, 
where $M_2$ stands for the 
mass of both $\psi_2$ and $\psi^{\pm}$. The 
effective degrees of freedom $g_{eff}$ in Eq. (\ref{sigma_eff}) is given by
 \begin{equation}
 g_{eff} = g1+ g_2 (1+\Delta)^{3/2} exp(-x\Delta) + g_3 (1+\Delta)^{3/2} exp(-x\Delta)\,.
\end{equation}  

\begin{figure}[htb!]
$$
\includegraphics[height=4.8cm]{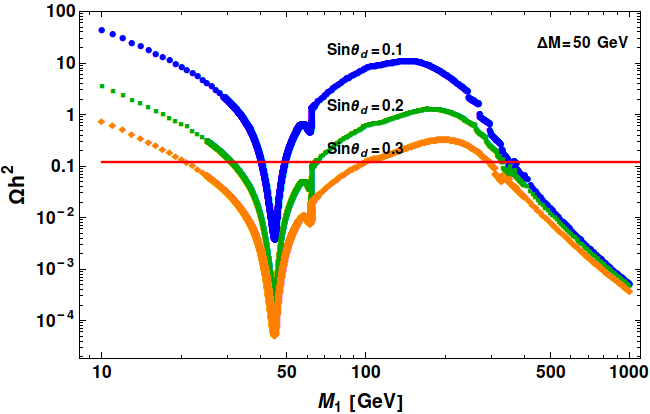}
\includegraphics[height=4.8cm]{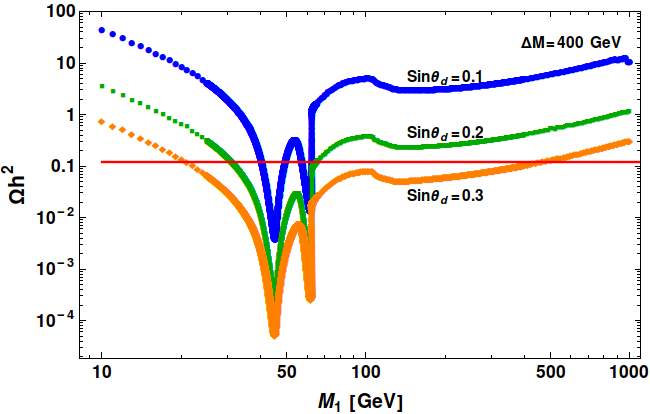}
$$
\caption{Relic density vs DM mass $M_1$ (in GeV) for different choices of $\sin\theta_d = \{0.1, 0.2,0.3\}$ with $\Delta M = 50$ GeV [left] 
(corresponding to $Y=\{0.02,0.04,0.058\}$  with blue, green, orange respectively) 
 and $\Delta M = 400$ GeV [right] (corresponding to $Y=\{0.16,0.32,0.46\}$ with Blue, Green, Orange respectively). Horizontal lines define the correct relic density.}
\label{fig:relic1}
\end{figure}

As it turns out from the above discussion, the dark-sector phenomenology in our set-up is mainly dictated by 
three parameters $\sin\theta_d, M_{1}$ and $\Delta M$. However we will keep on changing $\sin\theta_d$ and/or 
$\Delta M$ dependence with $Y$ wherever required using Eq.(\ref{theta-d}). In the following 
we use the code {\tt MicrOmegas}~\cite{Belanger:2008sj} 
to find the allowed region of correct relic abundance for our DM candidate $\psi_1$ 
satisfying  PLANCK constraints~\cite{planck,Ade:2015xua}, 
\begin{equation}
0.1175\leq \Omega_{\rm DM} h^2 \leq 0.1219 \,.
\label{eq:wmap.region} 
\end{equation}

In Fig. \ref{fig:relic1} we plot relic density versus DM mass $M_1$ for different choices of $\sin\theta_d = 0.1, 0.2$ 
and 0.3 (represented by blue, green and orange dotted lines respectively) while keeping the mass difference 
$\Delta M$ fixed at 50 GeV in the left panel and at $\Delta M = 400$ GeV in the right panel. The choice of various 
$\sin\theta_d$ can be translated into different values of $Y$ as well, through Eq. (\ref{theta-d}) since $\Delta M$ is kept fixed. 
Then it is equivalent to say that the blue, green and orange dotted lines in the left panel ($\Delta M$ = 50 GeV) represent 
$Y$= 0.02, 0.04, 0.058 respectively. In a similar way, the blue, green and orange dotted lines in 
the right panel ($\Delta M$ = 400 GeV) represent $Y$ = 0.16, 0.32, 0.46 respectively.  We infer that as 
the mixing increases or in other words $Y$ increases ($\Delta M$ is fixed), the doublet component starts 
to dominate (see Eq. (\ref{theta-d})) and hence give larger cross-section which leads to a smaller DM abundance for a particular $M_1$. 
The second important point to note is the presence of $Z$ resonance at $M_1=M_Z/2 \sim 45$ GeV and a Higgs resonance 
at $M_1=M_H/2\sim 63$ GeV  where relic density drops sharply due to increase 
in annihilation cross-section.
 We can also see that with larger $\Delta M$, $i.e.$ with larger $Y$ 
(as $\sin\theta_d$ is fixed) in the right hand side, 
the Higgs resonance 
is more prominent for obvious reasons.  Relic density for these chosen 
parameters are satisfied across 
the $Z$ resonance window and $H$ resonance window (more prominent for larger $\Delta M$ on the right panel). 
For small $\Delta M=50$ GeV (left panel of Fig.~\ref{fig:relic1}), relic density drops beyond DM mass of 300 GeV. 
This is due to co-annihilation channels start contributing $\overline{\psi_2}\psi_1 \to SM$ or $\overline{\psi^+}\psi_1 \to SM$ and 
we find that the relic density is satisfied for DM mass $\sim$ 400 GeV. This is however not seen in the right panel where we have larger $\Delta M$. 
This is because with the large mass gap, co-annihilation doesn't contribute significantly due to Boltzmann suppression 
for DM masses upto TeV.  That is why with larger $\Delta M$ (right 
panel of Fig.~\ref{fig:relic1}), 
there is no point for DM mass beyond 100 GeV associated with smaller 
$\sin \theta_d$ values like 0.1, 0.2, where relic density constraint is 
satisfied. 
With larger $\sin \theta_d=0.3$ one can satisfy relic density without the aid of co-annihilation for $M_1 \sim$ 500 GeV. We also note
a small drop in relic density on the right panel in particular, when $WW$ and 
$ZZ$ channels open up for annihilation.

\begin{figure}[thb!]
$$
\includegraphics[height=4.8cm]{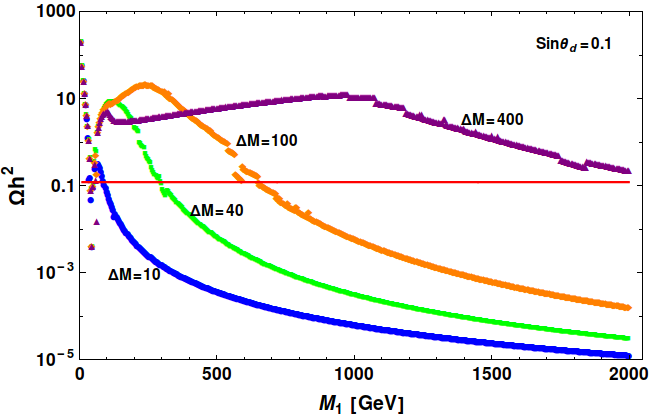}
\includegraphics[height=4.8cm]{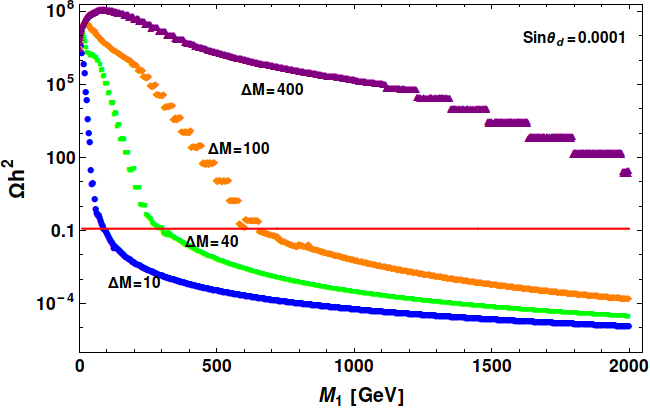}
$$
\caption{ Left: $\Omega h^2$ versus Dark matter mass $M_1$ (in GeV) for 
$\sin\theta_d=0.1$  with different choices of $\Delta M= \{10, 40, 100, 400\}$ 
GeV 
described by $\{$ blue, green, orange, purple respectively$\}$. Right: Same as 
left panel but with different $\sin\theta_d=0.0001$. Horizontal lines indicate 
correct relic density. }
\label{fig:Omega-m3}
\end{figure}

In order to show the effect of co-annihilations more closely, we draw 
Fig. \ref{fig:Omega-m3}, where one can see the $\Delta M$ dependency 
on relic density for a specific choice of mixing angle. In the left panel we choose $\sin \theta_d=0.1$ and that in the right 
panel for $\sin \theta_d=0.0001$. The slices with constant $\Delta M$ is shown for $\Delta M=\{10,40,100,400\}~{\rm GeV}$ in blue, green, orange, purple
lines respectively. We note here, that with larger $\Delta M$,
annihilation cross-section increases due to enhancement in Yukawa coupling ($Y 
\propto \Delta M$ as $\sin\theta_d$ is fixed). 
However, co-annihilation 
decreases due to increase in $\Delta M$ as $\sigma \propto e^{-\Delta M/M_1}$ 
specifically for a particular DM mass. Hence the larger is $\Delta M$ the 
smaller is co-annihilation and the larger is the relic density. This is clearly 
visible in both the panels of Fig. \ref{fig:Omega-m3}. In particular, when 
$\sin\theta_d$ is small, the effect of co-annihilation is pronounced as 
contribution from annihilation cross section is less dominant. This is the case 
shown in the right panel of Fig. \ref{fig:Omega-m3}. 
Hence the bigger is $\Delta M$, the larger is the required DM mass to satisfy relic density for a given 
mixing angle $\sin \theta_d $.  This is evident from the plot with $\Delta M = 
400$ GeV.

 For extremely small mixing angle, say $\sin \theta_d=0.0001$ (shown on 
the right panel of Fig. \ref{fig:Omega-m3}), 
the annihilation of $\bar{\psi}_1 \psi_1, \bar{\psi}_1 \psi_2 \to 
SM $ particles are highly suppressed. As a result the dominant contribution to relic density arises 
from $\psi_2 \psi^\pm , \psi^+ \psi^- \to$ SM particles.  This is 
an interesting consequence 
of our model.  In this case we get a lower limit of the singlet-doublet
mixing angle by assuming that the $\psi_2, \psi^\pm$ particles decay to $\psi_1$ before the latter freezes out 
from the thermal bath~\cite{Bhattacharya:2015qpa}. If the mass splitting between $\psi^-$ and $\psi_1$ is larger than $W^\pm$-boson mass, 
then $\psi^-$ decay preferably occurs through the two body process: $\psi^- \to \psi_1 + W^-$. However, if the mass splitting between $\psi^-$ 
and $\psi_1$ is less than $W^\pm$ boson mass, then $\psi^-$ decays through three body process, say $\psi^- \to \psi_1 \ell^- \overline{\nu_\ell}$. 
For the latter case, we get a stronger lower bound on the mixing angle than for two body decay. For the above mentioned channel, the three body decay width 
of $\psi^-$ is given by \cite{Bhattacharya:2015qpa}:
\begin{equation}\label{psi-decay}
\Gamma = \frac{ G_F^2 sin^2\theta_d}{24 \pi^3} M_2^5  I
\end{equation}
where $G_F$ is the Fermi coupling constant and $I$ is given as:
\begin{equation}\label{decay-rate}
I=\frac{1}{4}\lambda^{1/2}(1,a^2,b^2) F_1(a,b) + 6 F_2 (a,b)\ln  \left(\frac{2a}{1+a^2-b^2-\lambda^{1/2}(1,a^2,b^2)} \right) \,. 
\end{equation}
In the above Equation $F_1 (a,b)$ and $F_2 (a,b)$ are two polynomials of $a=M_1/M_2$ and $b=m_\ell/M_2$, where $m_\ell$ is the 
charged lepton mass. Up to ${\cal O}(b^2)$, these two polynomials are given by
\begin{eqnarray}
F_1 (a,b) &=& \left( a^6-2a^5-7a^4(1+b^2)+10a^3(b^2-2)+a^2(12b^2-7)+(3b^2-1)\right)\nonumber\\
F_2 (a,b) &=&  \left(a^5+a^4+a^3(1-2b^2)\right)\,.
\end{eqnarray} 
In Eq. (\ref{decay-rate}), $\lambda^{1/2}=\sqrt{1+a^4+b^4-2a^2-2b^2-2a^2b^2}$ defines the phase space. In the limit $b=m_\ell/M_2 
\to 1-a=\Delta M/M_2$, $\lambda^{1/2}$ goes to zero and hence $I\to 0$. The life time of $\psi^-$ is then given by 
$\tau \equiv \Gamma^{-1}$. Now to compare the life time of $\psi^-$ with DM freeze out epoch, we assume that the freeze out 
temperature of DM is $T_f= M_1/ 20$. Since the DM freezes out during radiation dominated era, the corresponding time of DM freeze-out is given by :
\begin{equation}
t_f= 0.301 g_\star ^{-1/2} \frac{m_{\rm pl}} {T_f^2} \, ,
\end{equation}
where $g_\star$ is the effective massless degrees of freedom at a temperature $T_f$ and $m_{\rm pl}$ is the Planck mass. Demanding 
that $\psi^-$ should decay before the DM freezes out (i.e. $\tau \lesssim t$) we get 
\begin{equation}\label{theta_constraint}
\sin \theta_d \gtrsim 1.1789 \times 10^{-5} \, \, \left(\frac{1.375\times
  10^{-5}} {I} \right)^{1/2} \left(  \frac{200 \rm GeV }{M_2}
\right)^{5/2} \left( \frac{g_\star}{106.75} \right)^{1/4} \left (
  \frac{M_1} {180 \rm GeV}\right)\,.
\end{equation}
Notice that the lower bound on the mixing angle depends on the mass of $\psi^-$ 
and $\psi_1$.

\begin{figure}[thb!]
$$
\includegraphics[height=4.8cm]{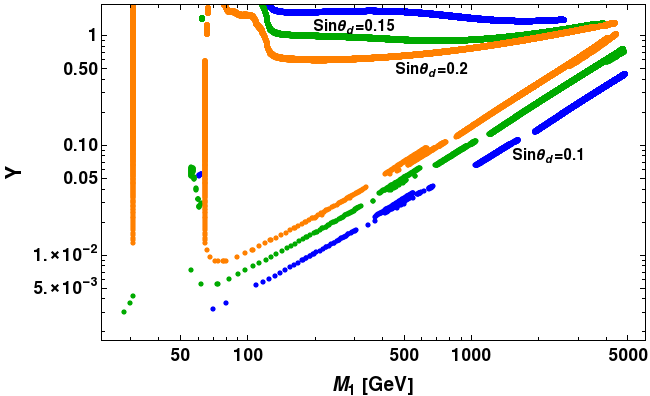}
\includegraphics[height=4.8cm]{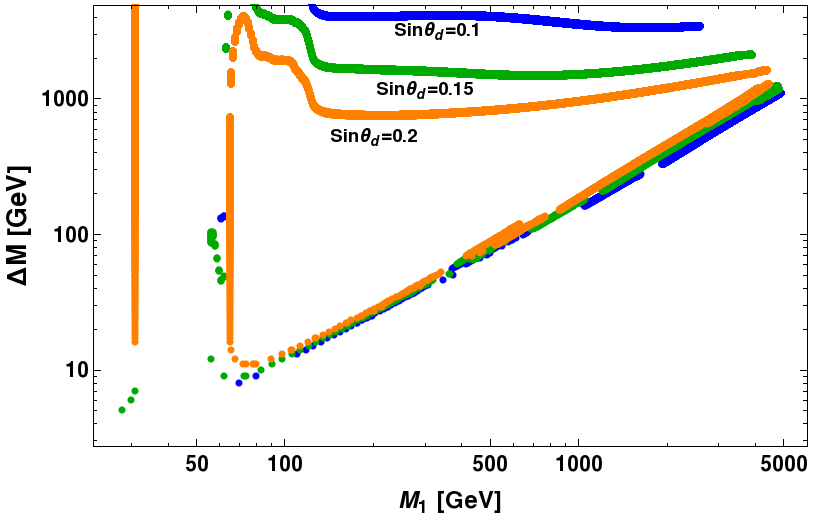}
$$
\caption{Left: $Y$ versus $M_1$ (in GeV) for correct relic density (Eq. \ref{eq:wmap.region}). $\sin \theta_d=0.1,~0.2,~0.15$
(blue, green and orange respectively) has been chosen, while $\Delta M$ vary arbitrarily. Right: Same plot in $M_1-\Delta M$ plane.} 
\label{fig:Y-M1}
\end{figure}

In Fig. \ref{fig:Y-M1} (left), we plot $Y$ versus $M_1$ to produce correct relic density with $\sin\theta_d=\{0.1, ~0.15,~0.2\}$ 
(blue, orange, green respectively). In order to be consistent with Eq. (\ref{theta-d}), $\Delta M$ 
has to be adjusted accordingly. It points out a relatively wide DM mass range satisfy the relic density constraint. 
  Main features that emerge out of this figure are as follows: (i) 
Firstly, there exist a lower DM mass region where $Z$ and $H$
resonances occur. Relic density is easily satisfied in this region for all possible moderate choices of $\sin\theta_d$, independent
of $Y$ or $\Delta M$ as is seen on the left hand side vertical lines (in both the plots). For large $\sin\theta_d$ this is more
prominent as both $Z$ and $H$ mediation is enhanced with larger mixing. (ii) The other point is to note that there are two 
regions for each $\sin\theta_d$ value which satisfy relic density; one at the below, where $Y$ (on the left) and $\Delta M$
(on the right panel) increase with larger DM mass to satisfy relic density. This region is dominantly contributed
from co-annihilations as the small $Y$ is not enough to produce annihilations required for relic density. While there 
is a second region with larger $Y$ (on left) and larger $\Delta M$ (on right), more insensitive to DM mass, where 
relic density is satisfied by appropriate annihilation cross-section, not aided by co-annihilations. Both of these 
regions (annihilation and co-annihilation domination) meet at some large DM mass $\sim$ 5000 GeV, more clearly visible
from the right panel plot. Points above the `correct annihilation lines' (for specific $\sin\theta_d$) provide more 
than required annihilation and hence those are under abundant regions. Similarly just below those, the annihilation
will not be enough to produce correct density and hence are over abundant regions. Points below the correct co-annihilation
regions produce more co-annihilations than required and hence depict under 
abundant regions.

\begin{figure}[thb!]
$$
\includegraphics[height=4.8cm]{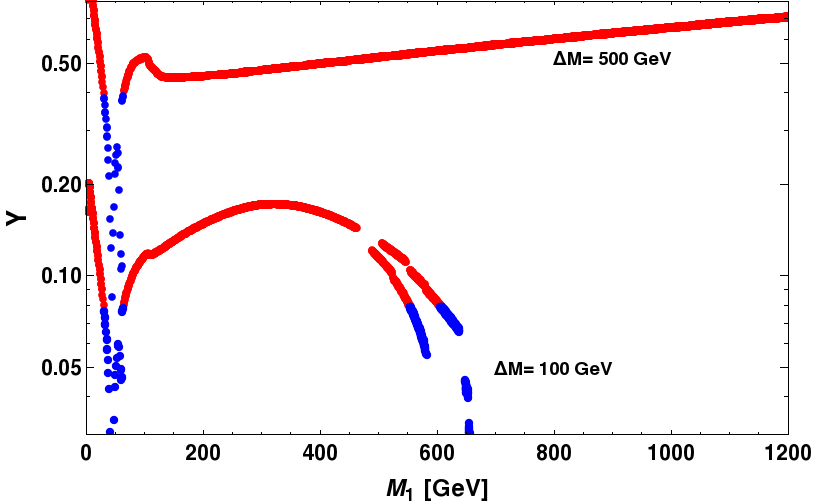}
\includegraphics[height=4.8cm]{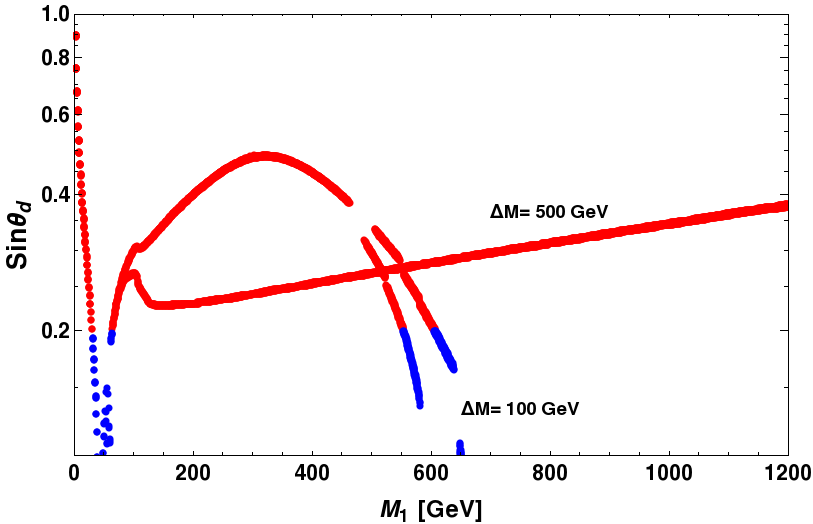}
$$
\caption{ Left: $Y$ versus $M_1$ (in GeV) for correct relic density (within the 
range given by Eq. (\ref{eq:wmap.region})) with fixed
$\Delta M$ at 100, 500 GeV, while $\sin \theta_d$ is allowed to vary. 
Right: Same plot in $M_1-\sin \theta_d$ plane. In both panels, blue dots 
are the allowed points for small $\sin\theta_d$ to satisfy the $Y, ~\Delta M$ 
abundance via Eq.~(\ref{theta-d}).} 
\label{fig:Y-M2}
\end{figure}

 The other possible correlation in this model for correct relic density 
can be drawn between DM mass ($M_1$) 
and the mixing angle ($\sin\theta_d$) for fixed $\Delta M$. This is shown in Fig.~\ref{fig:Y-M2} both in 
 $M_1-Y$ plane (on the left) or in $M_1-\sin\theta_d$ plane (on the right). For illustration, we choose two 
widely different values of mass difference: $\Delta M = 100$ GeV and $\Delta M = 
500$ GeV. 
This is clearly understood that with larger $\Delta M$, a larger $Y$ is favored for a specific DM mass in order to 
satisfy the correct relic abundance. With $\Delta M$ = 100 GeV we also note that $Y$ drops substantially 
around $M_1 \sim$ 500 GeV. This is because around this value, co-annihilation process starts contributing and 
hence it requires a further drop in $Y$ (in terms of mixing angle $\theta_d$) to obtain right relic density which is 
clearly visible in the right side of Fig. \ref{fig:Y-M2} as well. 
 Here we would like to draw the attention that the right relic density 
line has a split when co-annihilation starts
dominating. This is due to the fact that there are two different co-annihilations that occur here with $\psi_2$ and $\psi^{\pm}$. 
There exist a slight mass difference between these particles and the DM mass is adjusted to either of them to 
effectively co-annihilate and produce right relic density. For $\Delta M$ = 
500 GeV, this 
phenomena of co-annihilation occurs at a very large DM mass and can't be seen from the plot. 
Resonance drops both in $Y-M_1$  and $\sin\theta_d - M_1$ plots can be observed for 
$M_1 \sim M_H/2$ and $M_1 \sim M_Z/2$.  We also note that beyond 
$\sin\theta_d\ge 0.2$ as shown by the 
red points in Fig. \ref{fig:Y-M2} break small $\theta_d$ limit as has been 
assumed in 
Eq.~(\ref{theta-d}) and hence discarded within this approximation.
\begin{figure}[thb!]
$$
\includegraphics[height=6.5cm]{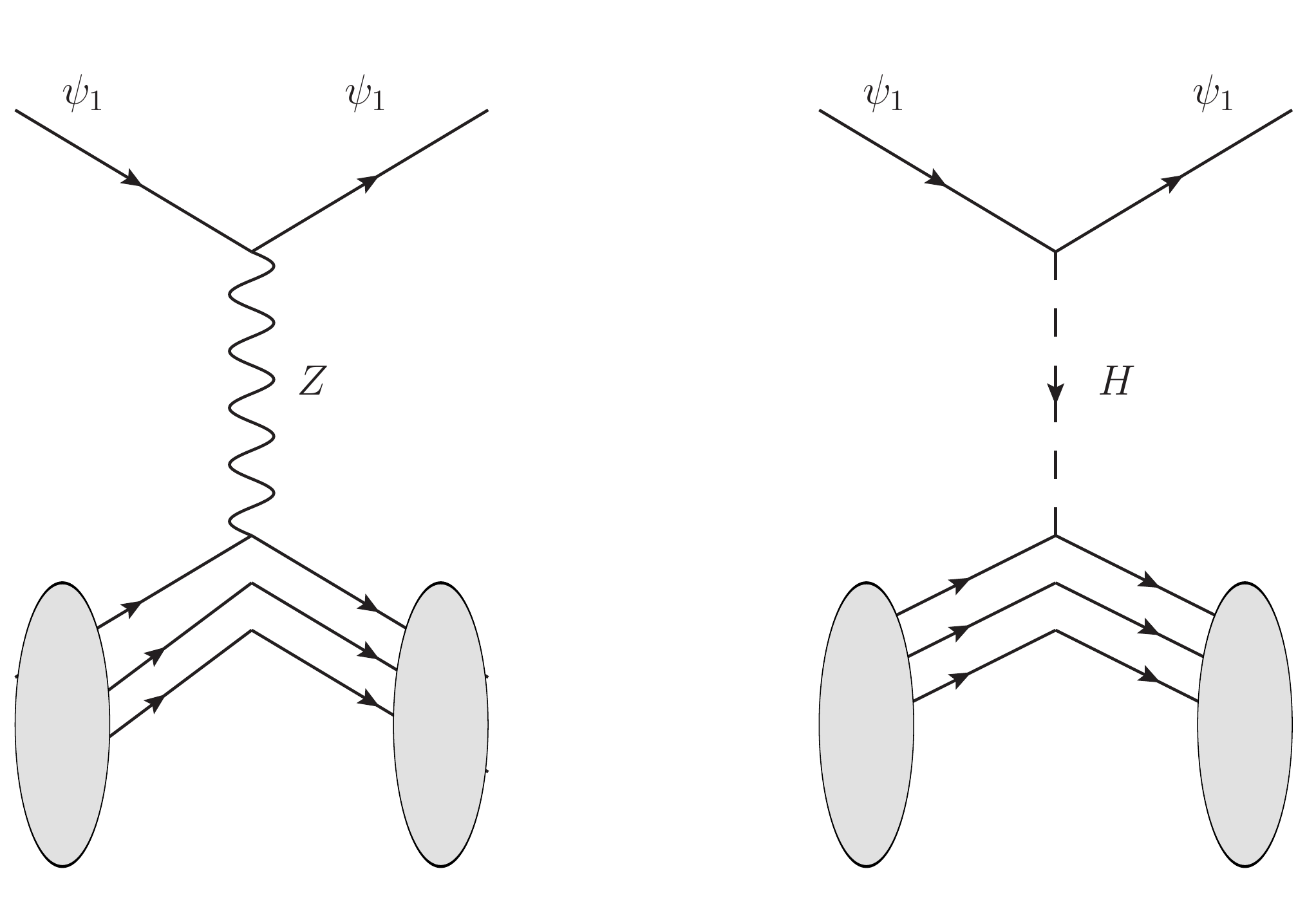}
$$
\caption{Feynman diagrams for DM to interact with Nucleon.} 
\label{fig:DD-fd}
\end{figure}

\begin{figure}[htb!]
$$
\includegraphics[height=4.8cm]{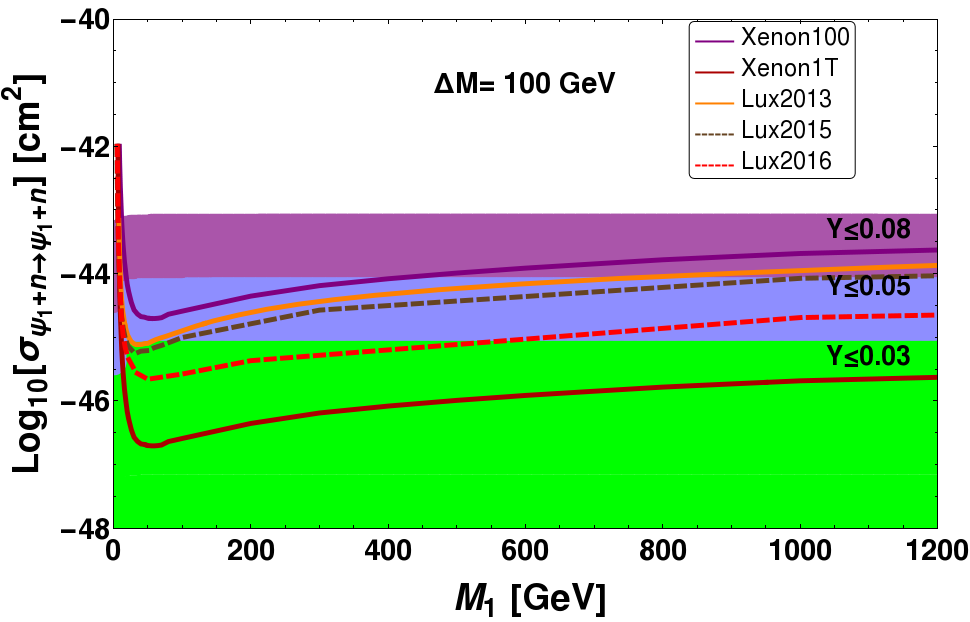}
\includegraphics[height=4.8cm]{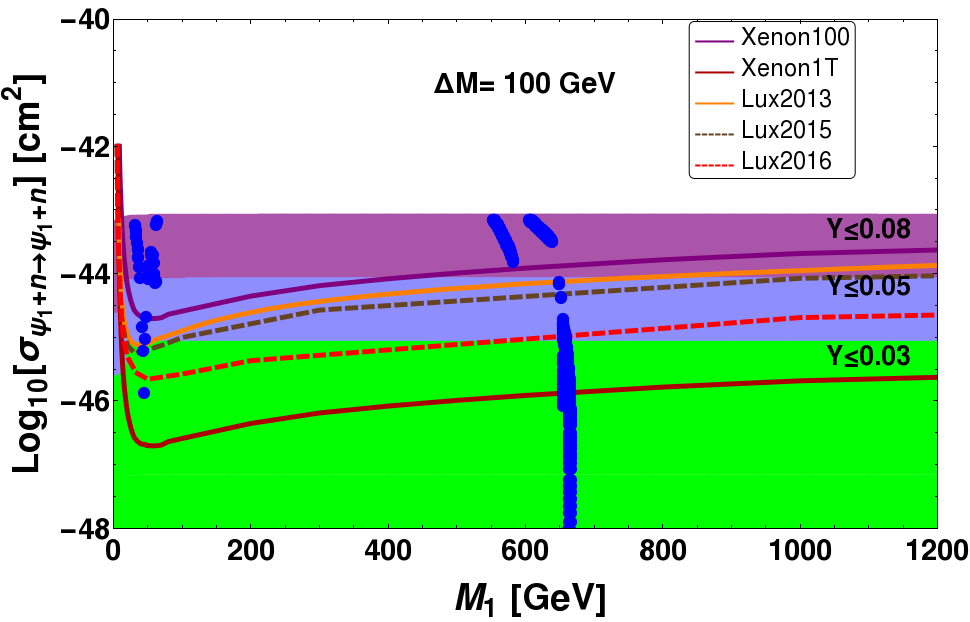}
$$
$$
\includegraphics[height=4.8cm]{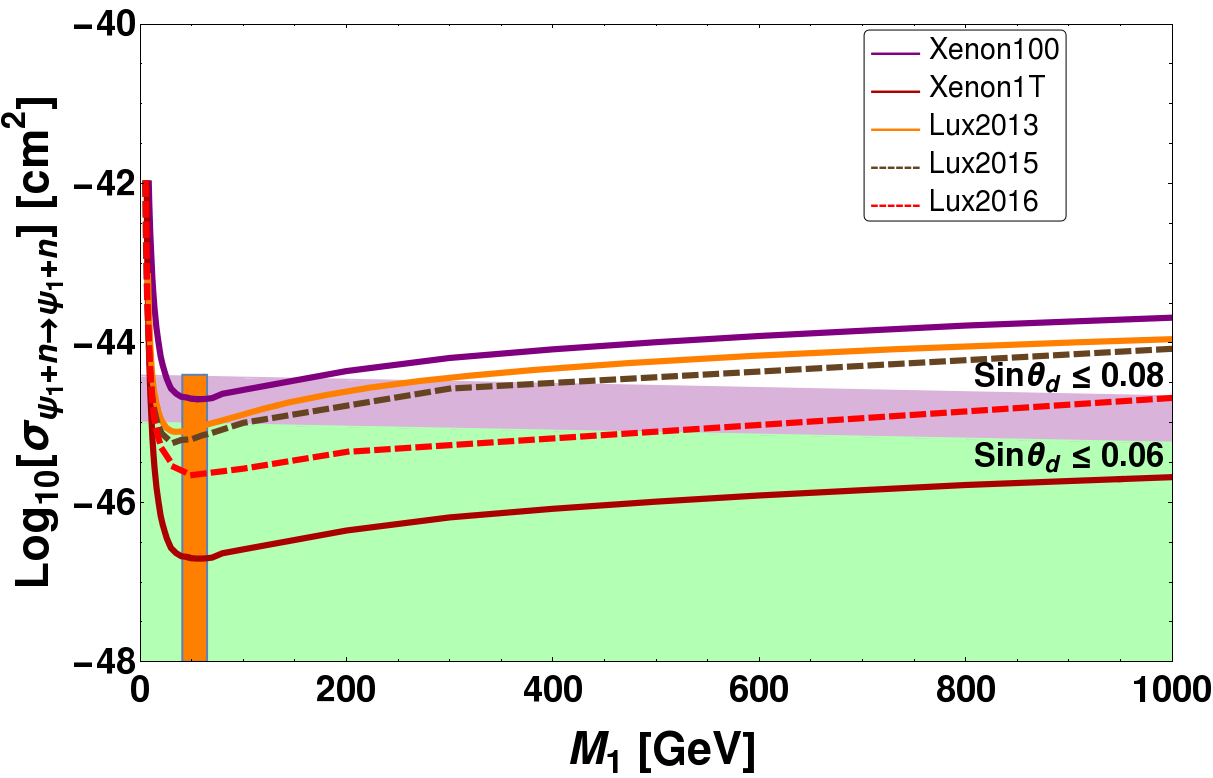}
\includegraphics[height=4.8cm]{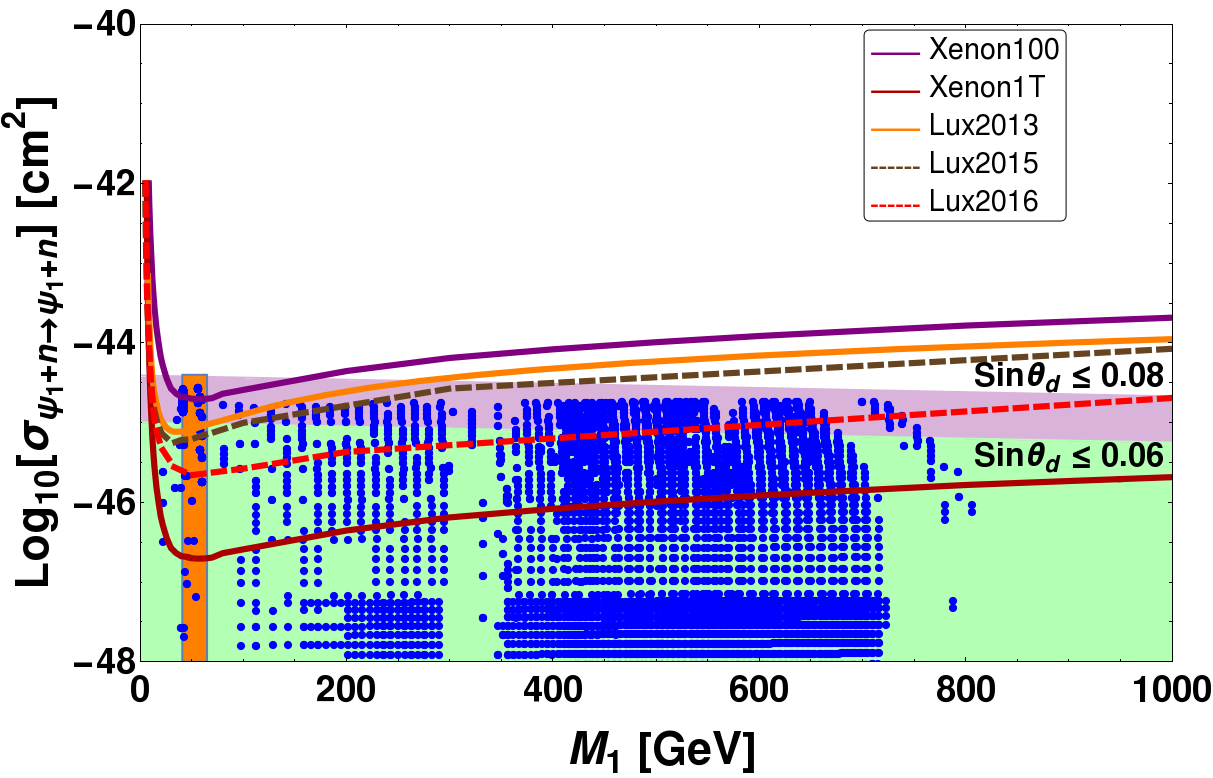}
$$
\caption{Spin independent direct search cross-section as a function of DM mass. 
Upper Left Panel: Different Y ranges are indicated $Y:\{0.001-0.03\}$ 
(green), $Y:\{0.03-0.05\}$ (blue) and $Y:\{0.05-0.08\}$ (purple). $\Delta M= 100$ GeV is 
used for the scan. Upper Right Panel: Same as left, additional blue dots represent points which 
satisfy relic density constraint. Lower left panel: Allowed ranges of $\sin \theta_d \le 0.06,0.08$ 
(light green and lilac regions respectively) are shown. Here $\Delta M$ varies 
arbitrarily upto 1.1 TeV. Lower right panel is same as the lower left panel having blue dots representative of
points which satisfy relic density constraint. The resonance region
is separately indicated in orange. Constraints from Xenon100, Lux 2013, 2015, 2016 
data and predictions of Xenon1T are presented.}
\label{fig:DD}
\end{figure}

Non-observation of DMs in direct search experiments tend to put a stringent bound on WIMP DM parameter space. Direct 
search interactions for $\psi_1$ has two different channels, through $Z$ and $H$ mediation as shown in
Fig.~\ref{fig:DD-fd}, where the one through $Z$ mediation dominates over $H$ mediated interaction because of 
$SU(2)$ gauge coupling. The cross-section per nucleon for $Z$ mediation is given by~\cite{ Goodman:1984dc,Essig:2007az}
\begin{equation}\label{DM-nucleon-Z}
\sigma_{\rm SI}^Z = \frac{1}{\pi A^2 }\mu_r^2 |\mathcal{M}|^2   
\end{equation}
where $\mu_r=M_1 m_n/(M_1 + m_n) \approx m_n$ is the reduced mass, $m_n$ is the mass of nucleon (proton or neutron),
$A$ is the mass number of the target nucleus and $\mathcal{M}$ is the amplitude for $Z$-mediated DM-nucleon cross-section
\begin{equation}\label{Z-mediated-process}
\mathcal{M}= \sqrt{2} G_F [Z (f_p/f_n) + (A-Z)] f_n \sin^2 \theta_d\,,
\end{equation}
$f_p$ and $f_n$ are the interaction strengths of DM with proton and neutron respectively and $Z$ is the atomic number
of the target nucleus. Using $f_n \simeq 1/3$ ~\cite{Koch:1982pu,Gasser:1990ap,Pavan:2001wz,Bottino:2008mf}, we obtain direct search cross-section per 
nucleon to be 
\begin{equation}
\sigma_{\rm SI}^Z \simeq 3.75 \times 10^{-39} {\rm cm}^2 \sin^4 \theta_d\,.
\end{equation}
Higgs mediated cross-section depends on can be written as 

\begin{equation}\label{scalar_mediated_crossssection}
\sigma_{\rm SI}^h=\frac{1}{\pi A^2}\mu_r^2  \left[ Z f_p + (A-Z)f_n \right]^2
\end{equation} 
where the effective interaction strengths of DM with proton and neutron are given by:
\begin{equation}
\label{f-values}
f_{p,n} = \sum_{q=u,d,s}f_{Tq}^{(p.n)} \alpha_q \frac{m_{(p,n)}}{m_q} + \frac{2}{27} f_{TG}^{(p,n)}\sum_{q=c,t,b} \alpha_q \frac{m_{p.n}}{m_q}
\end{equation}
with 
\begin{equation}
\label{alpha-value}
\alpha_q = \frac{ Y\sin 2\theta_d}{M_h^2} \left( \frac{m_q}{v}\right) \,.
\end{equation}

We compute the direct search cross-section with both diagrams using {\tt MicrOmegas}~\cite{Belanger:2008sj}. It turns out that
the most stringent constraint on the model and hence on the portal coupling $Y$ ($\lesssim \sin 2\theta_d \Delta M/(2v))$ comes
from the direct search of DM from updated LUX data \cite{Akerib:2016vxi} as demonstrated in Fig. \ref{fig:DD}. 
 We show the correct region of direct search allowed parameter space in 
two ways: in upper panel we choose a specific 
$\Delta M$ and vary $\sin\theta_d$ to evaluate spin independent direct search cross-section and show the constraints in terms of $Y$.
On the upper right panel, we also show the relic density allowed points through blue dots for this particular choice of $\Delta M$. 
In the bottom panel of Fig.~\ref{fig:DD}, instead of choosing a specific $\Delta M$, we vary it arbitrarily upto 1.1 TeV and 
point out the direct search constraints in terms of mixing angle $\sin\theta_d$. On the right bottom panel, we also show 
the relic density allowed points through blue dots. Restricting direct search 
cross-section to experimental limit actually
puts a stringent bound on mixing angle $\sin\theta_d$ to tame Z-mediated diagram in particular. We see that the bound from LUX,
constraints the coupling: $Y \sim 0.03$ for DM masses $\gtrsim 600$ GeV (green regions in the upper panel of Fig.~\ref{fig:DD}).
The Yukawa coupling needs to be even smaller for small DM mass for example, $M_1 \simeq 200$ GeV. The resonance region is exempted
from this constraint for obvious reasons. The annihilation cross-section is enhanced due to s-channel contribution and to tame it
to right relic density, one needs much smaller values of mixing angle, which sharply drops the direct search cross-section. Though
large couplings are allowed by correct relic density, they are highly disfavored by the direct DM search at terrestrial experiments.
 From the top right figure, we also see that correct relic density points 
for a specific $\Delta M$ lies in the vicinity of a 
specific DM mass $\sim$ 700 GeV where co-annihilation plays the crucial role for correct 
relic density and that doesn't contribute to direct search cross-section at all, so that the blue points yield very small direct
search cross-sections. This can easily be extended for other choices of $\Delta M$, where there exist a specific DM mass at which
co-annihilation plays a crucial role to yield right relic density, which doesn't contribute to direct search and thus can have very
small direct search cross-section as is seen from the right bottom figure. Note 
also that direct search constraints are less 
dependent on $\Delta M$ as to the mixing angle, which plays otherwise a crucial role in the relic abundance of DM.  In bottom panel, 
we show the parameter space satisfied by relic density constraint for $\sin \theta_d\le0.08,0.06$ (lilac and green 
regions respectively) to direct search constraints. The direct search tightly constraints the mixing angle to $\sin \theta_d \le 0.08$,
allowing DM masses as heavy as 900 GeV. Tighter constraint in mixing angle, for example, $\sin \theta_d \le 0.06$, allows smaller DM 
mass $\ge 500$ GeV as can be seen from the cross-over of LUX constraint with relic density allowed parameter space. 

In summary, the dark sector phenomenology with the inclusion of vector-like fermions provides a simple extension to SM, with a
rich phenomenology with a large region of allowed parameter space from relic density constraints. Direct search on the other 
hand constrains the mixing to a small value $\le 0.08$, allowing co-annihilation to play a dominant part to keep the model alive.
We will focus on the correlations to non-zero $\theta_{13}$ and DM in the following section with the results obtained from above
analysis.   Note that the $U(1)$ symmetry being global, its spontaneous 
breaking would lead to potentially 
dangerous Goldstone boson ($G$ = Im$\phi$). The problem however can be evaded by gauging the symmetry. Additionally 
if we assume the corresponding gauge boson to be sufficiently heavy, its existence will not modify our 
results of the dark matter phenomenology. Another way out is to provide tiny mass to the Goldstone by introducing 
an explicit symmetry breaking term in the Lagrangian. In this case however the most significant coupling of the Goldstone 
with Higgs appears through $\lambda_{12} \phi^{\dagger} \phi H^{\dagger} H$ coupling. Hence it contributes 
(considering $m_G \ll m_h/2$) to 
the invisible decay of the SM Higgs boson \cite{Joshipura:1992hp}, $\Gamma_{h \rightarrow G ~G} \sim 
\frac{1}{32 \pi} [m_h^3/\langle \phi \rangle ]\sin ^2\alpha$, where $\alpha$ signifies the mixing between the states 
$(H, \phi)$ and the physical Higgs fields $(h, H')$ resulting ($H'$ is the heavy Higgs) from non-zero $\lambda_{12}$. 
In the limit of $\lambda_{12}$ to zero, $\alpha$ vanishes. Using the present limit on the branching ratio of Higgs 
invisible decay \cite{Aad:2015txa, Khachatryan:2016whc} , the coupling $\lambda_{12}$ (involved in the 
definition of mixing angle $\alpha$) is expected to be small ($\ll 1$). If we assume a very small value of 
$\lambda_{12}, ~\sim 10^{-8}$ or even smaller, then it can be shown that the Goldstone can never be in thermal 
equilibrium \cite{Burgess:2000yq} and hence they can not contribute to the primordial abundance through freeze 
out mechanism \footnote{In this case, the other option could be \cite{Frigerio:2011in} the freeze-in mechanism 
\cite{Hall:2009bx}. It requires a detailed study and is at present beyond the scope of 
current analysis.} and we may basically ignore its presence for our purpose.

\begin{figure}[thb!]
$$
\includegraphics[height=7.5cm]{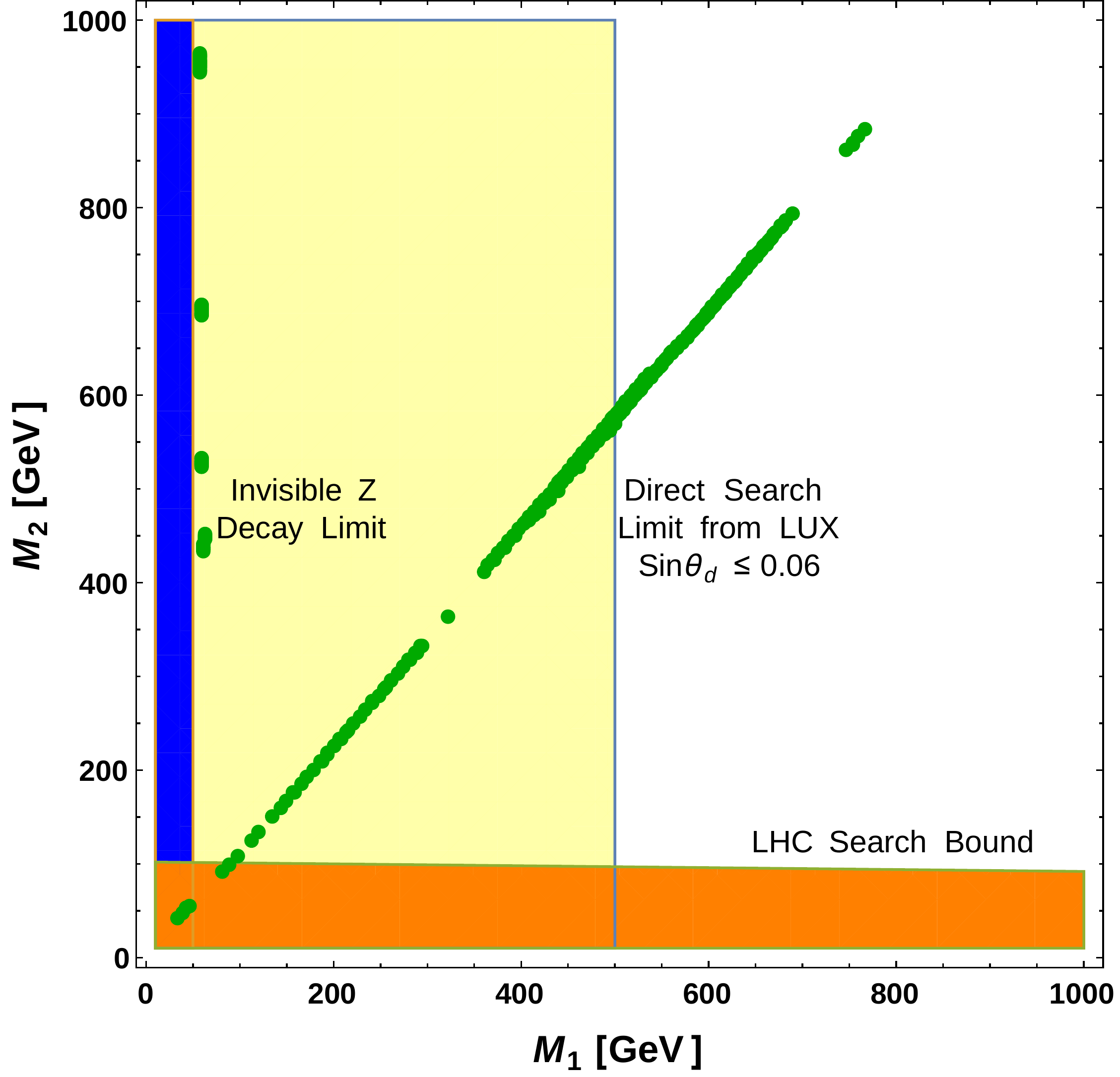}
$$
\caption{Summary of all constraints in $M_1-M_2$ parameter space from relic density 
 with $\sin\theta_d = 0.06$ (green dots), direct search (Yellow region is forbidden by updated LUX 
with $\sin\theta_d \sim 0.06 $), invisible Z-decay 
(blue region is forbidden) and collider (LHC) search limit (orange region is disallowed
with an over estimation for $\sin\theta_d = 0.1$).}
\label{fig:summary}
\end{figure}

We can now put together all the constraints for a specific choice of $\sin \theta_d=0.06$ into the plane of $M_1-M_2$ to 
show the allowed parameter space of the model. This is what we have done in Fig. \ref{fig:summary} following 
\begin{align*}
\label{eq:constraints}
\nonumber
\rm{Inv ~Z~ decay}: M_1< \frac{M_z}{2}\sim 45 ~GeV & \to \sin \theta_d \lesssim 0.00125 \\
\nonumber
\rm{Inv~ H ~decay}: M_1< \frac{M_h}{2}\sim 63~ GeV & \to \sin \theta_d \lesssim 0.1\\
\nonumber
\rm{Relic ~Density}: M_2 \lesssim M_1+100~ GeV & ~{\rm for}~ \sin \theta_d \lesssim 0.1\\
\nonumber
\rm{Direct ~Search}: M_1 \ge 500 ~GeV & ~{\rm for}~ \sin \theta_d \sim 0.06\\
\rm{Collider~ Bound}: M_2 \simeq M^{\pm} \ge 101 ~GeV & ~{\rm for}~ \sin \theta_d \sim 0.06.
\end{align*}

 We choose $\sin\theta_d=0.06$ as a reference value  as it satisfies all of the constraints discussed here. We see that a 
sizable part of the DM parameter space is allowed shown by the green dotted points, excepting for the direct search 
bound shown by yellow band, a blue band disfavored by the Invisible $Z$ decay and orange band disfavored by 
direct collider search data \cite{collider}. One should also note here that if we choose a smaller $\sin\theta_d$ to
illustrate the case, a larger DM mass region is allowed by direct search constraint.
Green dotted points show relic density allowed regions 
of the model in $M_1-M_2$ plane. We note here that for $\sin\theta_d < 0.1$, only co-annihilation can provide with right relic 
density, hence is independent of the choices $\sin\theta_d \sim 0.1$ or $\sim 
0.06$ as has been chosen in Fig. \ref{fig:summary}.

\section{Correlation between Dark and Neutrino Sectors}\label{DM_Neutrino}

As stated before, our description of the DM sector is composed of a vector like $SU(2)_L$ 
doublet and a neutral singlet fermions which interact with the SM sector via Eq. (\ref{lagrangian}). We have seen 
in the previous section the importance of the effective coupling $Y$ in determining the mixing between 
the singlet and doublet components of DM (see Eq. (\ref{theta-d})). This mixing in turn plays the crucial role in realizing 
the correct relic density as well as involved in the direct search cross section (see Eqs. (\ref{eq:omega}) and
 (\ref{DM-nucleon-Z},~\ref{scalar_mediated_crossssection})). 
Note that this effective coupling $Y$ is generated from the vev of the flavon $\phi$ through $Y = \epsilon^n$, 
where the $n$ is the unknown $U(1)$ charge assigned to $\phi$. However this vev alone does not 
appear separately in our dark matter analysis. On the other hand, we have noted earlier the involvement 
of $\epsilon$ parameter in the neutrino phenomenology, in particular in producing $\theta_{13}$ in the 
correct ballpark. So we observe that the allowed value of nonzero $\theta_{13}$ and the Higgs portal 
coupling of a vector like dark matter can indeed be obtainable from a $U(1)$ flavor extension of the SM.
In this section we aim to fix the charge $n$ from combining the results of neutrino as 
well as the dark matter analyses. This complementarity between the neutrino and the DM sector will be 
clear as we proceed below in summarizing constraints on  $\epsilon$ and $Y$ obtained from neutrino and 
DM analyses respectively.  

Section \ref{neutrino_section} was devoted to neutrino phenomenology, where we 
have discussed four different cases. In case A, 
we find that the parameter $\epsilon$ is clearly determined to be within the range $ 0.328- 0.413$ in order to keep 
$\sin\theta_{13}$ in agreement with experimental data (see Fig. \ref{fig:s}). In cases B and C however, 
this correlation between $\epsilon$ and $\theta_{13}$ is not that transparent as it depends also on 
the CP phase $\delta$. Combining all the phenomenological constraints  ($e.g.$ on $\sum_i m_{\nu_i}$),
we have provided the range of $\epsilon$ in Table \ref{tab:3} and \ref{tab:4} for cases B and C respectively. The range of 
$\epsilon$ corresponding to case D is similar to case A. On the other hand, the information on $Y$ is 
embedded in the relic density and direct detection cross section.
\begin{figure}[thb]
$$
\includegraphics[height=4.8cm]{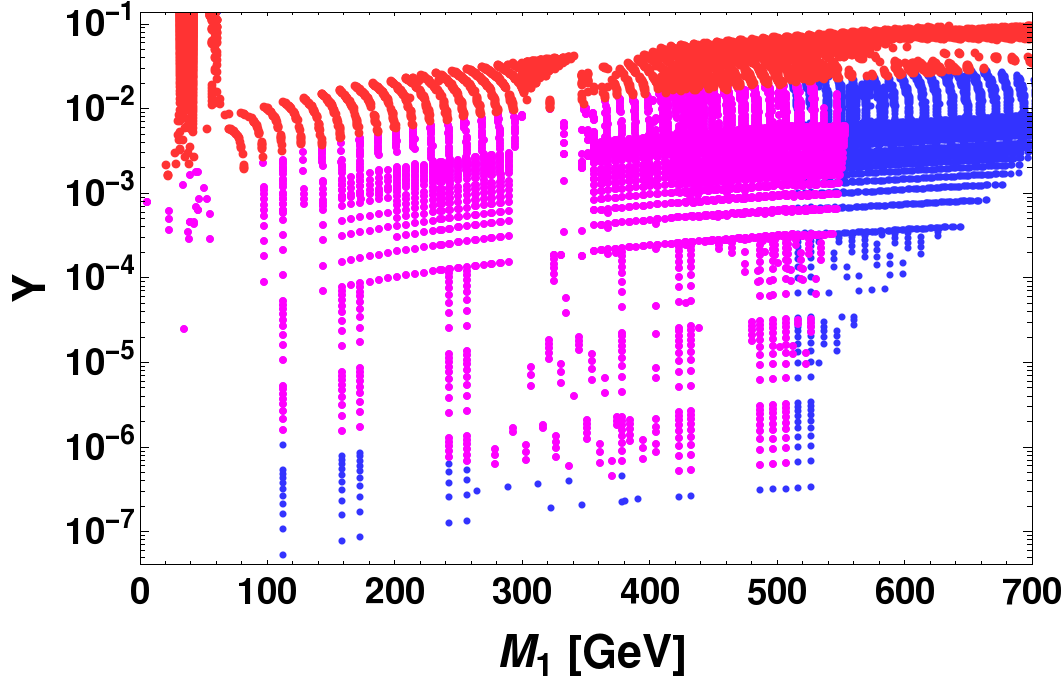}
\includegraphics[height=4.8cm]{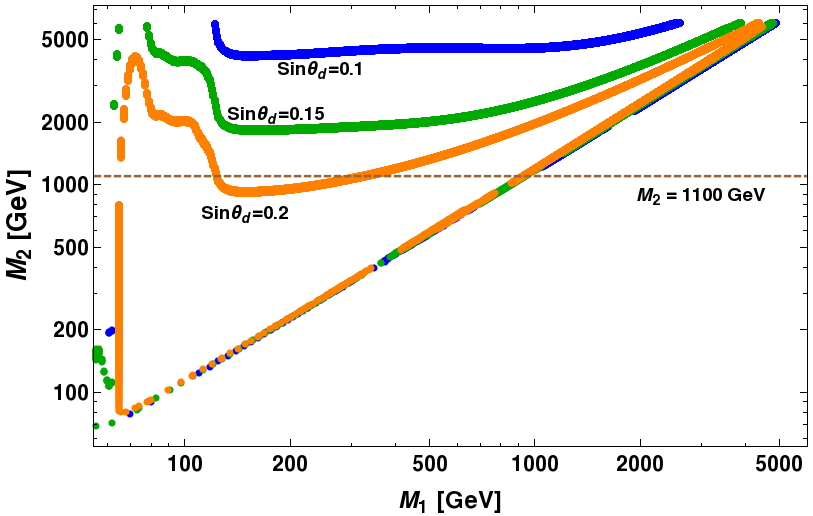}
$$
\caption{Left Panel: $Y$ vs $M_1$ scatter plot for correct relic density (Eq. 
\ref{eq:wmap.region}). Here
 $\sin\theta_d$ ($10^{-6}$-0.2)  and $\Delta M$ (1-1100 GeV) varies 
simultaneously. The top red points are disallowed by Lux 2016 
direct search constraint. Both magenta and blue dots simultaneously 
satisfies relic density and Lux 2016 direct search constraints. The magenta 
dots additionally satisfies  the condition $\Delta M < M_W$. Right Panel: $M_1$ 
versus $M_2$ (in GeV)
for correct relic density. $\sin \theta_d=0.1,~0.2,~0.15$
(blue, green and orange respectively) has been chosen, while $\Delta M$ varies. 
The left panel is consistent with this plot upto $M_2$ = 1.1 TeV as marked by 
the 
horizontal dashed line in this plot. }
\label{fig:YvsM}
\end{figure}

In the left upper panel of Fig. \ref{fig:DD}, we plot 
the direct search cross-section against 
dark matter mass $M_1$ for a fixed choice of $\Delta M = 100$ GeV. In this plot, we indicate regions allowed by direct search experimental 
limits. Since each point in the region allowed by direct search 
correspond to a specific relic density, once we incorporate both the relic density and 
direct search limit by LUX 2016, we 
find the allowed region is narrowed down as shown in the right upper panel of Fig. \ref{fig:DD} (indicated by 
blue patch).

 Similarly the left lower panel (left and right) of Fig. 
\ref{fig:DD} shows the allowed (by both relic density and LUX 2016) 
region of parameter space where variation of $M_2$ is restricted up to 1.1 TeV with $\sin\theta_d \leq 0.2$. We find that an 
uper limit on $\sin\theta_d$ is prevailing from this plot. Combining relic density constraint and direct search limits, we find the 
allowed region indicated by blue dots in the right lower panel of Fig. \ref{fig:DD}. In order to obtain limits on $Y$ while 
$\Delta M$ and $\sin\theta_d$ are varied, we have provided a scatter plot of $Y$ versus $M_1$ in 
Fig. \ref{fig:YvsM}. In producing this plot, we have varied $M_2$ (up to 1.1 TeV), $10^{-7} < \sin\theta_d < 0.2$. 
Here red dots correspond to those points which are disallowed by LUX 2016 even if these satisfy the 
relic density constraint. The blue patch indicates the region allowed by both the relic density and 
LUX 2016 data having $\Delta M > m_W$. For $\Delta M < m_W$, we use a lower limit on $\sin\theta_d$ 
obtained from Eq. (\ref{theta_constraint}). Hence the points in magenta satisfy the above $\sin\theta_d$ constraint and represent the allowed 
region by relic density and direct search limits. From this plot we can clearly see the upper limit 
of $Y$ is almost 0.03 while the lower limit of it can be very small, $\sim 10^{-7}$. Note that the $Y$ region limited by the choice of upper value of $M_2$ = 1.1 TeV is consistent with 
our earlier plot in Fig. \ref{fig:Y-M1} with fixed $\sin\theta_d$ values. For 
elaboration purpose,
we provide the figure in the right panel of 
Fig. \ref{fig:YvsM}, which is the same plot as Fig. \ref{fig:Y-M1} except 
that it is now plotted in terms of $M_2$ vs. $M_1$. The narrow patch 
for a fixed $\sin\theta_d$ becomes wider as we varied $\sin\theta_d$ as well. The horizontal dashed line indicates our consideration of keeping 
the variation of $M_2$ within 1.1 TeV.

\begin{figure}[thb]
$$
\includegraphics[height=3.6cm]{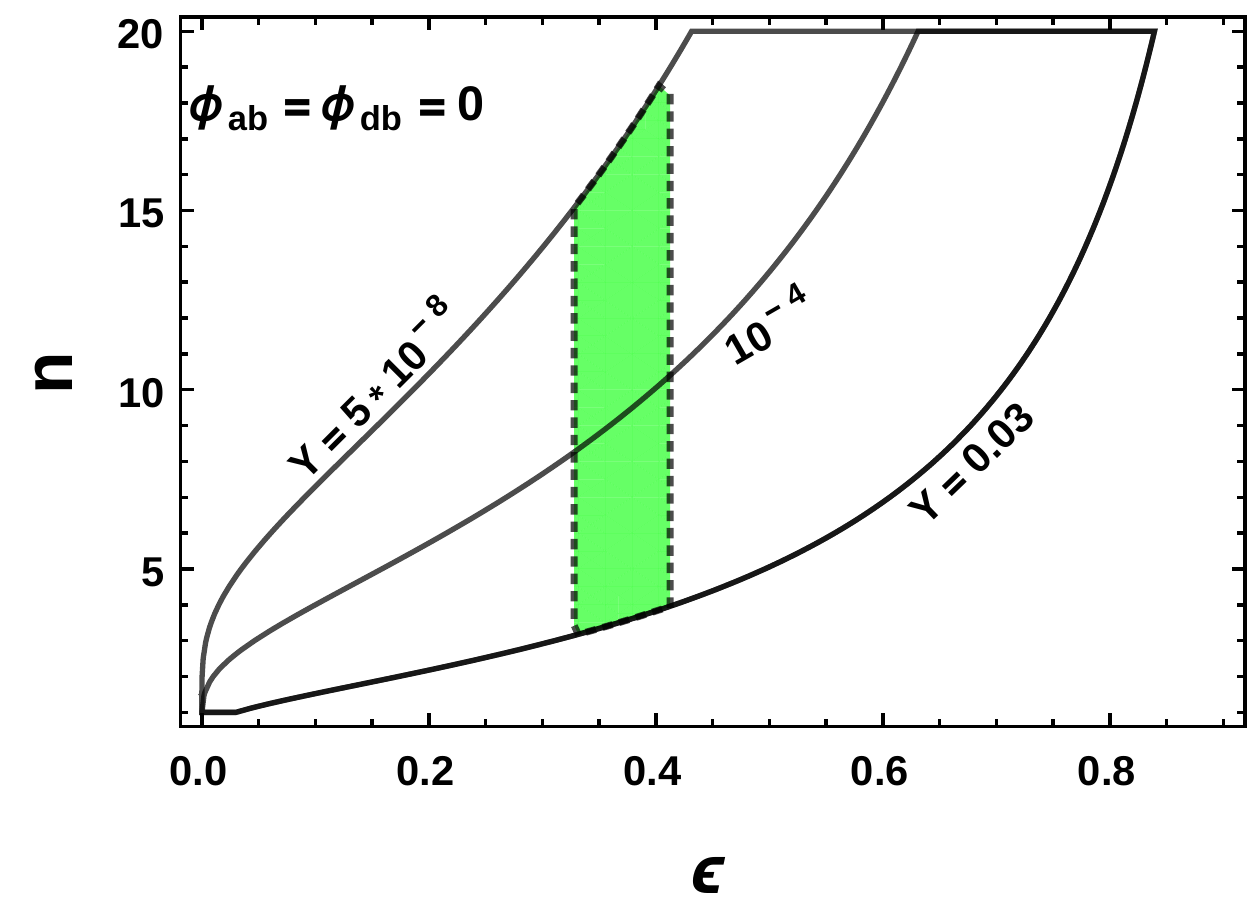}
\includegraphics[height=3.6cm]{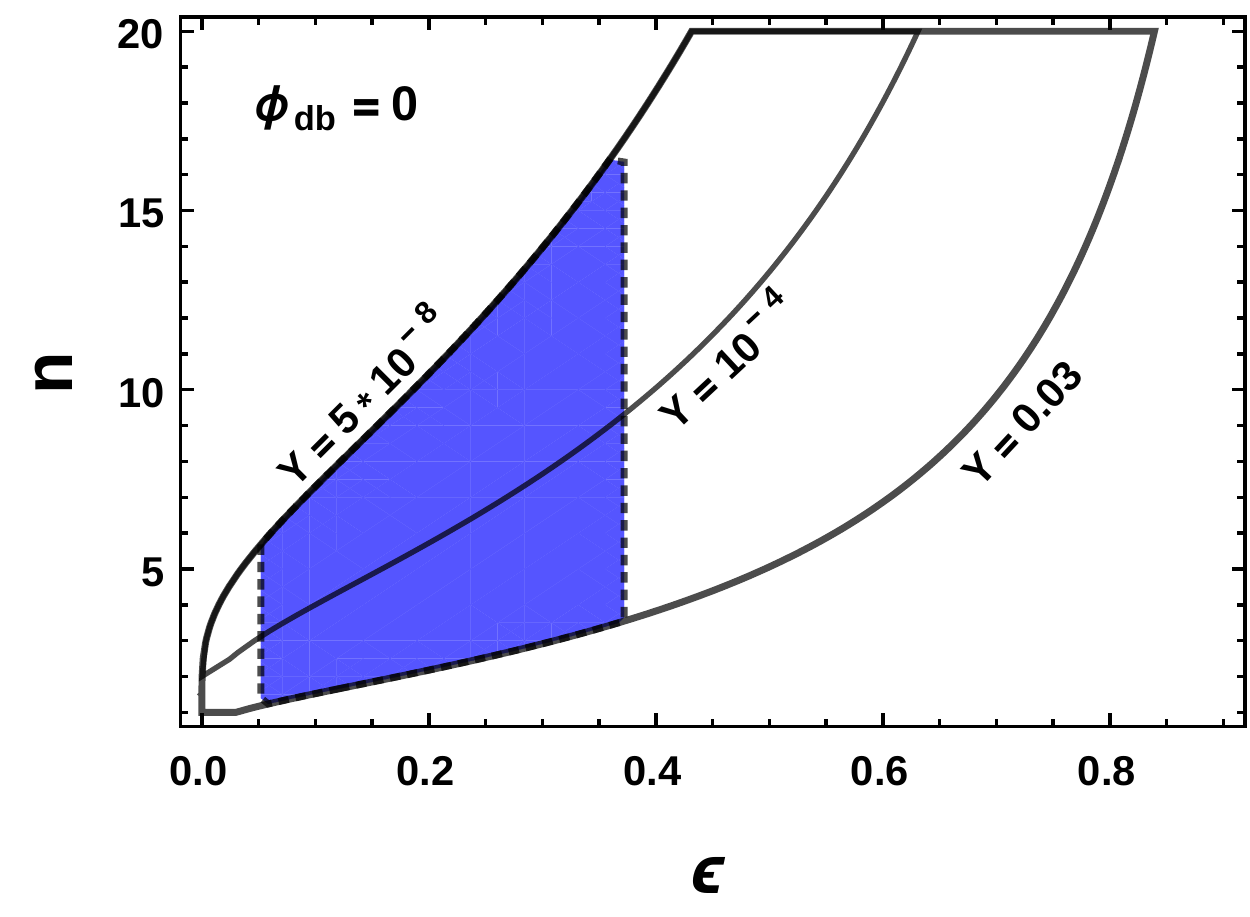}
\includegraphics[height=3.6cm]{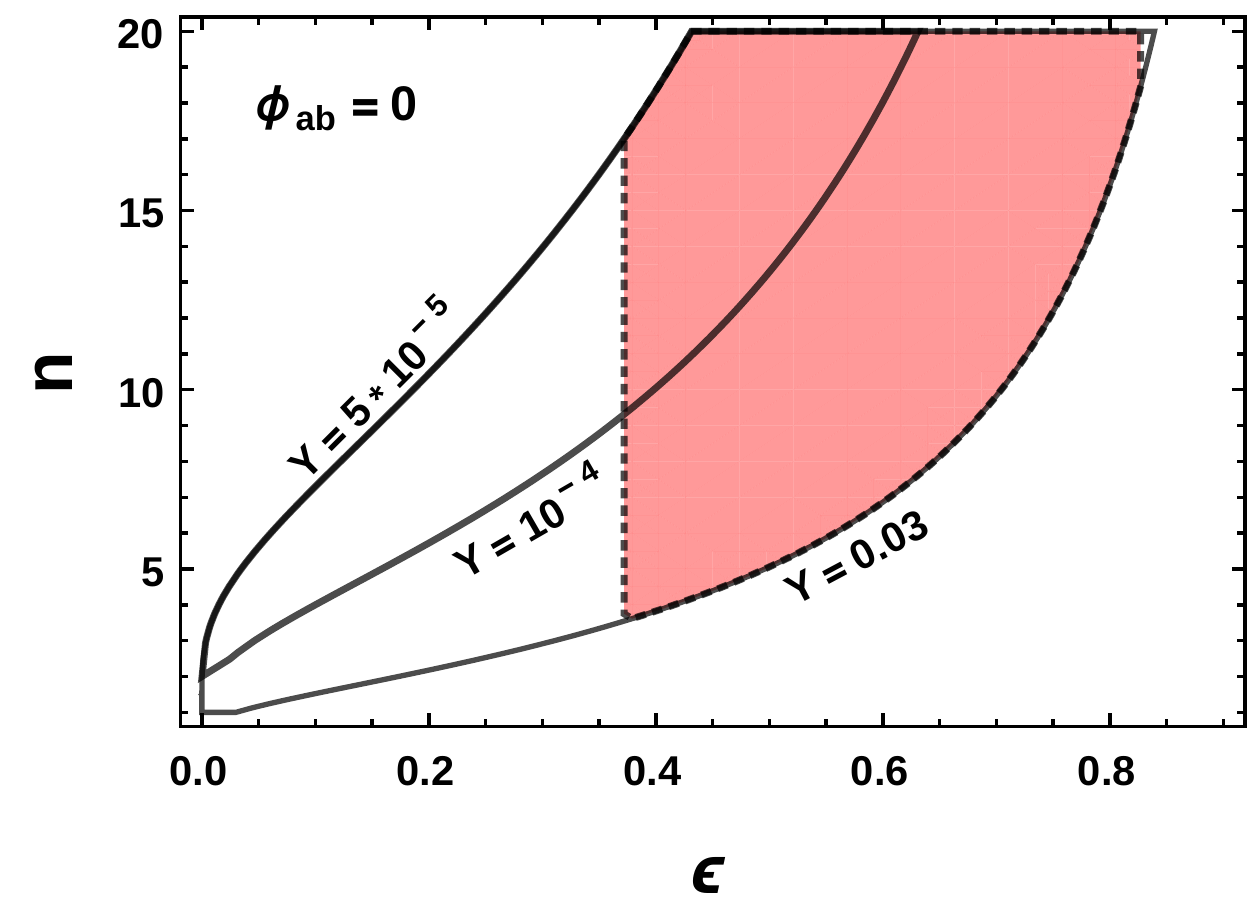}
$$
\caption{ $n$ vs $\epsilon$ to generate different values of $Y=\epsilon^n$ for (a) $\phi_{db}=\phi_{ab}=0$ (left), 
(b) $\phi_{db}=0$ (middle) and (c) $\phi_{ab}=0$ (right).}
\label{fig:epsilon}
\end{figure}

We summarize here these constraints 
on $\epsilon$ and $Y=\epsilon^n$ to determine the unknown flavor charge $n$ of the dark matter in our scenario. It 
is shown in Fig. \ref{fig:epsilon}. Colored patch in each plot corresponds to the allowed range of $\epsilon$ obtained 
in section \ref{neutrino_section} for Cases A(D), B and C. In the left-most panel of Fig. \ref{fig:epsilon}, we have shown the 
allowed values of $n$ where the CP-violating phases are taken to be zero corresponding to Case A.  As the direct 
search of DM restricts the $Y$ values to be $Y\lesssim 0.03$, we get $n\gtrsim 2$. Different contour lines with 
different $Y$ values are shown in the figure. A similar conclusion holds for the other case (Case D) with $\phi_{db} 
= \phi_{ab} = \beta$ . On the other hand, if $\phi_{ab}\neq \phi_{db}$ then a larger range of $n$ values 
are expected to be allowed. In particular, by setting $\phi_{db}=0$ and $\phi_{ab}\neq 0$ (as shown in middle panel 
of Fig. \ref{fig:epsilon}) we see that lower limit on $n$ starts from 1. On the other hand, if $\phi_{ab}=0$ 
and $\phi_{db}\neq 0$ (as shown in the right panel of Fig. \ref{fig:epsilon}) then $n$ can take values starting 
from 3. Thus we conclude that the non-zero values of phases introduce more uncertainty in specifying $n$.  
The future measurements of  Dirac CP phase $\delta$ and a more stringent constraints from Direct search 
experiments would reduce this uncertainty in $n$.

\section{Conclusions}\label{concl}

In this paper we have explored a $U(1)$ flavor extension of the SM in order to establish a possible correlation 
between the SM sector (more specifically neutrino sector) and the DM one, in particular between the reactor 
lepton mixing angle $\sin \theta_{13}$ and the interaction of dark matter with SM Higgs. To start with, 
we have considered a tri-bimaximal mixing  pattern ($i.e.$ with $\theta_{13} = 0$) for the lepton mixing 
matrix originated from a typical flavor structure of the neutrino mass matrix guided by the non-Abelian 
flavor symmetry, where the charged lepton mass matrix is found to be diagonal.  In its simplest version, 
we achieve the TBM structure of the neutrino mass matrix by assuming an $A_4 \times Z_3$ symmetry 
where the effective dimension six operators involving $A_4$ flavons contributes to Majorana masses for 
light neutrinos.  The symmetry forbids the usual dimension five operator. 
On the other hand, the dark sector consists of two vector-like fermions, one is a $SU(2)_L$ doublet and 
the other one is a SM gauge singlet. In addition we assume the existence of a $U(1)$ flavor symmetry 
under which the DM fields as well as two flavons, $\phi$ and $\eta$, are charged. It is interesting to note 
that with the vector-like fermions present in the dark sector, there exists a replica of SM Yukawa interaction 
in the dark sector which involves flavon $\phi$.  The $U(1)$ symmetry of the model was broken at a 
high scale by the vev of that flavon field $\phi$ to a remnant $\mathcal{Z}_2$ under which the dark sector 
particles are odd. As a result the lightest odd particles becomes a viable candidate of dark matter. 
Moreover, a higher dimensional operator involving $\phi$ and $\eta$ constitutes a correction to the 
TBM pattern of the neutrino mass matrix which leads to a non-zero value of $\sin \theta_{13}$. The 
involvement of $\phi$ ensures that $B-L$ breaking vev is also involved in this correction term. 
As a result we are able to show that the non-zero value of $\sin \theta_{13}$ is proportional to the 
Higgs portal coupling, $Y = (\phi/\Lambda)^n \equiv \epsilon^n$, of the dark matter which gives rise to 
correct relic density measured by WMAP and PLANCK and consistent with direct DM search bound from LUX. 
Finally it is interesting to note that $Y$, on one hand is related to the mixing in the 
neutrino sector, while it also crucially controlled by the mixing involved in the dark sector. We also find that the 
current allowed values of $\sin \theta_{13}$ indicates the $U(1)$ charge of DM $\gtrsim 1$ which can be probed 
at the future direct DM search experiments such as Xenon-1T. The next to lightest stable particle (NLSP) is a 
charged fermion which can be searched at the LHC~\cite{Arina:2012aj,Arina2}. In the limit of small $\sin\theta_d$, the 
NLSP can give rise to a displaced vertex at LHC, a rather unique signature of the model discussed in 
ref.~\cite{Bhattacharya:2015qpa}. We argue that this is a minimal extension to SM to accommodate DM and 
non-zero $\sin \theta_{13}$ by using a flavor symmetric approach.     
\acknowledgments{The work of SB is partially supported by DST INSPIRE grant no PHY/P/SUB/01 at IIT Guwahati. 
NS is partially supported by the Department of Science and Technology, Govt. of India under 
the financial Grant SR/FTP/PS-209/2011. }


\end{document}